\documentclass{aa}
\usepackage{natbib}
\bibpunct{(}{)}{;}{a}{}{,} 
\usepackage{graphicx}
\usepackage{txfonts}
\usepackage{hyperref}
\hypersetup{colorlinks, allcolors = {blue}}
\usepackage{xcolor}

\begin{document} 
   \title{Dynamics and Emission Properties of Flux Ropes from Two-Temperature GRMHD Simulations with Multiple Magnetic Loops}
\titlerunning{Flux ropes from GRMHD simulations}


   \author{Hong-Xuan Jiang,
          \inst{1} 
          Yosuke Mizuno, \inst{1,2,3}
          Indu K. Dihingia,
          \inst{1}
          Antonios Nathanail,
          \inst{4}
          Ziri Younsi,
          \inst{5}
          Christian M. Fromm
          \inst{6,3,7}
          }
    \authorrunning{Jiang et al.}

   \institute{Tsung-Dao Lee Institute, Shanghai Jiao Tong University, Shengrong Road 520, Shanghai, 201210, People's Republic of China\\
        \email{hongxuan\_jiang@sjtu.edu.cn, mizuno@sjtu.edu.cn}
         \and
             School of Physics and Astronomy, Shanghai Jiao Tong University, 800 Dongchuan Road, Shanghai, 200240, People’s Republic of China
        \and 
             Institut f\"ur Theoretische Physik, Goethe-Universit\"at Frankfurt, Max-von-Laue-Stra{\ss}e 1, D-60438 Frankfurt am Main, Germany
        \and 
             Research Center for Astronomy, Academy of Athens, Soranou Efessiou 4, GR-11527 Athens, Greece
        \and 
             Mullard Space Science Laboratory, University College London, Holmbury St. Mary, Dorking, Surrey, RH5 6NT, UK
        \and 
             Institut f\"ur Theoretische Physik und Astrophysik, Universit\"at W\"urzburg, Emil-Fischer-Str. 31, D-97074 W\"urzburg, Germany
        \and 
             Max-Planck-Institut f\"ur Radioastronomie, Auf dem H\"ugel 69, D-53121 Bonn, Germany
             }


 
  \abstract{
  Flux ropes erupting from the vicinity of the black hole are thought to be a potential model for the flares observed in Sgr\,A$^*$.}
  {In this study, we examine the radiative properties of flux ropes that emerged from the vicinity of the black hole.}
  {We have performed three-dimensional two-temperature General Relativistic Magnetohydrodynamic (GRMHD) simulations of magnetized accretion flows with alternating multiple magnetic loops, and General Relativistic Radiation Transfer (GRRT) calculations. In GRMHD simulations, two different sizes of initial magnetic loops are implemented.}
  {In the small loop case, magnetic dissipation leads to a weaker excitement of magneto-rotational instability inside the torus which generates a lower accretion rate compared to the large loop case. However, it makes more generation of flux ropes due to frequent reconnection by magnetic loops with different polarities. By calculating the thermal synchrotron emission, we found that the variability of light curves and emitting region are tightly related. At $230\,\rm GHz$ and higher frequency, the emission from the flux ropes is relatively stronger compared with the background, which is responsible for the filamentary structure in the images. At lower frequencies, e.g. $43\,\rm GHz$, emission comes from more extended regions, which have a less filamentary structure in the image.}
  {Our study shows self-consistent electron temperature models are essential for the calculation of thermal synchrotron radiation and the morphology of the GRRT images. Flux ropes contribute considerable emission at frequencies $\gtrsim 230\,\rm GHz$.}


   \maketitle


\section{Introduction}
The supermassive black hole (SMBH) in our galaxy (Sgr\,A$^*$) is one of the most mysterious objects. Despite many extensive research efforts to comprehend the physics of the central SMBH, we still have many open questions \citep[e.g.,][]{2002apa..book.....F,2010RvMP...82.3121G, 2012RAA....12..995M}. Recent EHT observations have made significant progress in understanding the basic properties of Sgr\,A$^*$ 
 \citep{event_horizon_telescope_collaboration_first_2022-4, event_horizon_telescope_collaboration_first_2022-3, event_horizon_telescope_collaboration_first_2022-2,event_horizon_telescope_collaboration_first_2022-1,event_horizon_telescope_collaboration_first_2022}. 
EHT observations of Sgr\,A$^*$ are tested with various astrophysical models. However, none of the models passed all observational constraints \citep{event_horizon_telescope_collaboration_first_2022}. Nonetheless, the EHT model constraint prefers magnetically arrested disk (MAD) models over Standard and Normal Evolution (SANE) flow models and other accretion flow models, including tilted disks \citep{Liska2018}, and stellar-wind-fed models \citep{Ressler2020}.
Both Sgr\,A$^*$ and M\,87 are categorized as low-luminosity Active Galactic Nuclei (LLAGNs) 
 \citep[e.g.,][]{Yuan2014, 2008ARA&A..46..475H}. 
Their accretion rates are far below the Eddington rate \citep{Ho2009, Marrone2007}. 
Many recent models have indicated that the accretion flows around the SMBH in M\,87 (M\,87$^*$) are possibly also in a MAD state \citep{EHTpaperV, Cruz-Osorio2021, Yuan2022}. 
However, although the accretion rates of the two SMBHs are similar, the physical properties of the accretion flow of Sgr\,A$^*$ are still not perfectly understood yet.

The observed hot spot in Sgr\,A$^*$ is another unsolved problem \citep{TheGRAVITYCollaboration2020, Lin2023, stad1106}. 
\cite{gravity_collaboration_detection_2018} detected orbital motions of a hot spot near the innermost stable circular orbit of Sgr\,A$^*$ at near-infrared (NIR) frequency. 
Their modeling shows a hot spot orbiting at a few gravitational radii away from the horizon, consistent with the observed flare activities \citep{TheGRAVITYCollaboration2020}. 
Flares in Sgr\,A$^*$ are observed at both sub-millimeter and NIR frequencies \citep[e.g.][]{Do20191,Do2019}. 
During the flares, the NIR flux increases several times from the magnitude of the quiescent state.
The NIR flux distribution suggests the total emission of Sgr\,A$^*$ is a combination of a quiescent state and sporadic flares \citep{Abuter2020}.
Polarization information of the hot spots of Sgr\,A$^*$ is provided by EHT-ALMA observation \citep{2022A&A...665L...6W}, which seems to be consistent with the vertical magnetic flux rope from the magnetically arrested region in accretion flows \citep[e.g.,][]{2020MNRAS.497.4999D,porth_flares_2021,Scepi2022}. In particular, \cite{2020MNRAS.497.4999D} has provided a clockwise rotating hot-spot motion in MAD flow around Sgr\,A$^*$ by GRMHD simulations. Furthermore, EHT observations suggest a small line of sight inclination angle \citep{EHTpaperV}, which is also recommended by \cite{VonFellenberg2023} through light curve fitting of the flares.
These MAD models suggest the flaring events are strongly related to the flux eruptions, which are located in the equatorial plane. However, polarimetric tomography suggests the orbital motion of the flare is in a low-inclination orbital plane \citep{Levis}. Plasmoid chains and flux ropes, typically generated in the jet sheath, exhibit a low inclination angle \citep{nathanail_plasmoid_2020,nathanail_magnetic_2022,stad1106, Aimar2023, Mellah2023}. Considering this, we propose that the plasmoid chain model offers a more plausible explanation for the observed flares.

The formation of the plasmoids in the accretion flow has been proven to be a plausible scenario in GRMHD simulations of magnetized accretion flows with multi-loop magnetic configurations \citep{nathanail_plasmoid_2020,stad1106}.
In these simulations, highly turbulent accretion flow and frequent reconnection events occur due to the merging magnetic loops. Plasmoid chains are also seen in the jet sheath in the 3D general relativistic particle-in-cell (GRPIC) simulations of \cite{Mellah2023}.
Due to the frequent reconnection from the merging magnetic loops, non-steady jets with low power are generated along with large numbers of plasmoids \cite[for detail, see][]{nathanail_plasmoid_2020,chashkina_grmhd_2021,nathanail_magnetic_2022,stad1106}. 
In our previous 2D GRMHD simulations of magnetized accretion flows with multiple magnetic loops \citep{stad1106}, plasmoid chains were generated mainly via the reconnection of a magnetic field with opposite polarity. In 2D, we only observe purely poloidal plasmoid chains due to underlying axisymmetric conditions. 
Recently \cite{nathanail_magnetic_2022} suggested that the plasmoid chains have rather three-dimensional structures. Accordingly, self-consistent evolution of toroidal component of the magnetic field becomes important, which requires 3D simulations. We expect to observe the filamentary structure, which may be instrumental in explaining the flaring activities. Unfortunately, it cannot be fully resolved in our simulations due to the resolution limit. Although the plasmoid chains in this work are not fully resolved, the two-temperature sub-grid models partially address this issue by providing a more accurate calculation of electron temperature. Consequently, we would be able to study the emission properties of these structures more precisely.

The presence of a mean-field dynamo in the accretion disk can change the initial magnetic configuration and make it easier to get alternating polarities of the magnetic field \citep{DelZanna2022, Mattia2020, Mattia2022}. However, previous simulations \citep{2021MNRAS.506..741M} have revealed that the alternating polarities generated within the magnetically arrested disk (MAD) do not significantly impact the formation of a strong jet. This limitation may arise from the finite numerical resolution, which weakens the amplification of the magnetic field by the dynamo process. Consequently, an additional dynamo term becomes necessary \citep[e.g.,][]{Mattia2020, Mattia2022,2024MNRAS.527.3018Z}. Even when the numerical resolution is sufficiently high, a persistent strong jet exists in the MAD regime \citep{Ripperda2021}. To achieve a sufficiently weak jet and promote frequent reconnection within the jet sheath, a multiple magnetic loop configuration remains essential, implying distinct polarities of plasma injection into the torus.

Previous simulations only examined accretion flow in single-fluid approximation, which can not produce the self-consistent radiative signatures \citep{stad1106}. Calculation of self-consistent synchrotron emissions from electrons in RIAFs requires a two-temperature framework \citep{Ressler2015}.
In our previous study, we used two-temperature GRMHD simulations of magnetized accretion flow with multiple loops to investigate electron temperature distributions, drawing on the electron heating prescriptions of \citep{Rowan2017, Kawazura2019}. Our findings revealed that the temperature distributions of plasmoids and the current sheet around black holes differ significantly when compared to the parameterized electron-to-ion temperature ratio prescription with two-temperature calculations \citep{stad1106}.
We have also shown that the flaring activity in Sgr~A$^*$ is likely related to the unique features of plasmoids formed in various magnetic field configurations. 
Following these findings, in this study, we extend our previous work.
We perform two-temperature 3D GRMHD simulations of magnetized accretion flows with multiple magnetic loops. 
In our investigations, the accretion flow formed from multi-loop configurations is mostly in the SANE regime. Although MAD models are more favored for the observations of Sgr~A$^*$, many SANE models also pass the EHT tests \cite{event_horizon_telescope_collaboration_first_2022}. Also, none of the MAD or SANE models used in the EHT analysis (with a single-loop magnetic field) can meet the light curve's variability requirement \citep{event_horizon_telescope_collaboration_first_2022}.
It suggests that neither single-loop SANE nor MAD is appropriate to explain all the observation features of Sgr~A$^*$. New accretion flow models are required to better understand the accretion flow of Sgr~A$^*$.  The results of multiple magnetic loops show unique features, including frequent magnetic reconnections with a relatively higher accreting magnetic flux than a normal SANE torus \citep{Fromm2022}. Therefore, we study the dynamics of the accretion flows from different sizes of magnetic loops to investigate possible applications for Sgr~A$^*$.
Subsequently, we try to understand the radiation properties of the accretion flows by performing GRRT calculations of thermal synchrotron emission.
Through a comparative analysis of the GRRT calculations and the emissivity distribution calculated from GRMHD data, we investigate the properties of emitting regions across various observing frequencies. In particular, we explore the time variability and the formation of hot spots. 

The rest of the paper is organized as follows: In Section 2, we describe our simulation setup, which includes the initial condition and magnetic field configurations used in this study. In Section 3, we discuss our findings from our GRMHD simulations and post-processed GRRT calculations. We provide our conclusions in Section 4.

\section{Numerical Method}

The numerical methods employed for the GRMHD simulation of this study are identical to those utilized in previous work \citep{2021MNRAS.506..741M, stad1106}. 
For the sake of completeness, we only describe the basic information here. 
The {\tt BHAC} code is used to perform the 3D GRMHD simulations, which solve the $3+1$ form of ideal GRMHD equations in geometric units ($G=M=c=1$ and $1/\sqrt{4}\pi$ is included in the magnetic field) \citep{Porth2017, Olivares2019}. The simulations are performed in Kerr spacetime, and we also neglect the self-gravity of matter.

The simulations are initiated from a rotation-supported hydrodynamic equilibrium torus following \citep{1976ApJ...207..962F}. To avoid repetitions, we do not show the explicit expressions for the initial quantities here. Instead, we only show the initial torus parameters, which are $r_{\rm in} = 20\, \rm r_{g}$ and $r_{\rm max} = 40\,\rm r_{\rm g}$, where $r_{\rm g}=GM/c^2$ is the gravitational radius and $M$ is the mass of the black hole. The constant adiabatic index $\Gamma=4/3$ is used \citep{Rezzolla2013}. In our study, we consider a black hole with a spin parameter of $a=0.9375$. The torus mass is assumed to be negligible compared to that of the black hole, resulting in a spacetime described by the static Kerr metric. 2\% of a random perturbation is applied to the gas pressure within the torus to excite the magneto-rotational instability (MRI). This instability generates turbulence in the accretion disk.

\begin{figure}
    \centering
         \includegraphics[width=.8\linewidth]{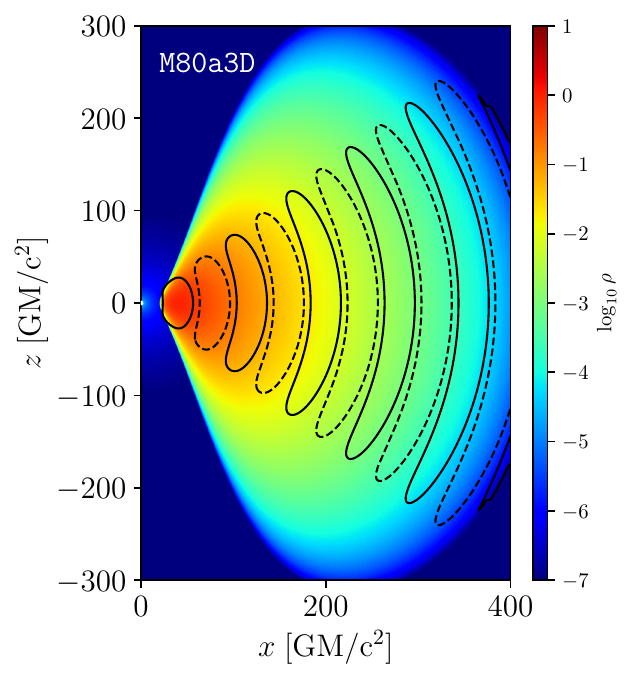}
    \caption{Initial torus profile of logarithmic density (color) and magnetic field configuration (black contours). Dashed contours represent positive polarity, while solid ones are the opposite.}
    \label{fig: initial condition}
\end{figure}

A pure poloidal magnetic field configuration is considered as an initial condition for the magnetic fields, which is set up with initial vector potential:
\begin{equation}
\begin{aligned}
    A_{\rm \phi}\propto& (\rho - 0.01)(r/r_{\rm in})^3\sin^3\theta \exp{(-r/400)}\\
    &\cos((N-1)\theta)\sin(2\pi(r-r_{\rm in})/\lambda),
\label{eq:A_phi}
\end{aligned}
\end{equation}
where $N=3$ is used in all cases, and $\lambda$ is a 
the wavelength of the magnetic configuration in the radial direction. 
The value of $A_{\rm \phi}$ is determined by setting the minimum of plasma $\beta_{\rm min}=100$, where $\beta \equiv p_{\rm g}/p_{\rm mag}$. $p_{\rm g}$ is the gas pressure, and $p_{\rm mag} = b^2/2$ is the magnetic pressure, where $b^2=b_{\mu}b^{\mu}$ and $b^{\mu}$ is the 4-magnetic field. Fig.~\ref{fig: initial condition} demonstrates the initial density distribution with the multiple poloidal magnetic loops (black lines). The solid and dashed contours represent different polarities of the poloidal magnetic loops. In Fig.~\ref{fig: initial condition}, we present the initial magnetic field configuration (black lines) of case {\tt M80a3D} with initial density distribution (color map) as an illustration of the alternating multi-loop magnetic configuration.
In the case {\tt M20a3D}, each loop length becomes shorter but the torus structure does not change.

In this work, we adopt the parameters of the two representative cases from our 2D simulations \citep{stad1106}, i.e., cases of {\tt M20a} and {\tt M80a} to perform 3D simulation, which are labeled as {\tt M20a3D} and {\tt M80a3D} respectively. We remind the readers that models {\tt M20a3D} and {\tt M80a3D} correspond to the cases with $\lambda=20$ and $\lambda=80$, respectively.

In this work, we adopt two electron heating prescriptions: turbulence heating and magnetic reconnection heating. 
The turbulence electron heating model was originally prescribed with damping of MHD turbulence, which is given by \citep{Kawazura2019}:
\begin{equation}
     f_{\rm e} = \frac{1}{1+Q_{\rm i}/Q_{\rm e}},
\end{equation}
where
\begin{equation}
     \frac{Q_{\rm i}}{Q_{\rm e}} = \frac{35}{1 + \left(\beta/15\right)^{-1.4}\exp\left(-0.1 T_{\rm e}/T_{\rm i}\right)}.
\end{equation}
Subsequently, for the reconnection heating prescription, we adopt a fitting function measured by PIC simulations, which is given by \citep{Rowan2017}:
\begin{equation}
     f_{\rm e} = \frac{1}{2} \exp\left[\frac{-(1-\beta/\beta_{\rm max})}{0.8+\sigma_{\rm h}^{0.5}}\right],
     \label{Eq: reconnection heating}
\end{equation}
where $\beta_{\rm max} = 1/4\sigma_{\rm h}$, $\sigma_{\rm h} = b^2/\rho h$ is magnetization as defined to the fluid specific enthalpy $h = 1 + \Gamma_{\rm g} p_{\rm g}/(\Gamma_{\rm g}-1)$.
Applying the turbulence and reconnection electron heating prescriptions simultaneously, we store the electron entropy from the two models separately and obtain the electron temperatures from the two heating models. The benefit of this method is that it avoids the non-linear effect in the GRMHD simulations and allows us to directly compare the different electron heating models in the same GRMHD data.

The outer boundary of the simulations is set at $r=2,500\,\rm r_g$. The simulations are performed in spherical Modified Kerr–Schild coordinates (see \cite{Porth2017}) with an effective grid resolution $384 \times 192 \times 192$, including the entire $2\pi$ azimuthal domain with three static mesh refinement levels. The simulation models {\tt M20a3D} and {\tt M80a3D} are evolved up to $t = 15,000\,\rm M$ and $20,000\,\rm M$, respectively. These simulation times are enough for both models to reach a quasi-steady state.

\begin{figure}
    \centering
         \includegraphics[width=\linewidth]{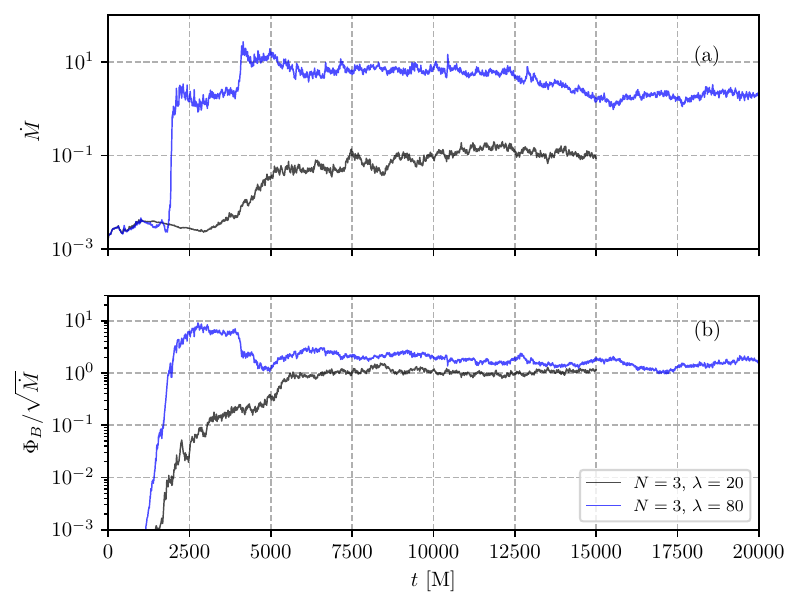}
    \caption{Time evolution of (a) mass-accretion rate ($\dot{M}$) and (b) magnetic flux rate ($\Phi_{\rm B}/\sqrt{\dot{M}}$) onto the event horizon in both {\tt M20a3D} (black) and {\tt M80a3D} (blue) cases.}
    \label{fig: Mdot} 
\end{figure}

To obtain the synthetic images from the GRMHD simulations, GRRT calculation is performed with {\tt BHOSS} code \citep{2012A&A...545A..13Y,2020IAUS..342....9Y} in post-process. 
The GRRT equations are solved along the geodesics and integrated through 3D GRMHD data. The resultant images, light curves, and spectra are obtained at a given viewing angle $i$ and observing frequency $\nu$ as seen by a distant observer.
In this work, all GRRT post-processed calculations use the same viewing angle at $i=30^{\circ}$, which is adapted from \cite{event_horizon_telescope_collaboration_first_2022}.
In the GRRT calculations, a threshold on the magnetization $\sigma=b^2/\rho$ is imposed to limit the emission regions in the high-magnetized domain because the high magnetization region in the GRMHD simulations may be affected by the floor treatment and thus unreliable. 
In this work, we only consider the region with $\sigma < 1$. We also adopted the fast light approximation, which neglects the propagation time of light. 
In this work, we use Sgr\,A$^*$ as the target for GRRT calculation. It has a mass $M_{\bullet}=4.5\times10^6\,\rm M_{\odot}$ and a distance $D_{\rm SgrA}=8.5\times10^3\,\rm pc$ from earth. The field of view (FOV) is $160\,\rm \mu as$ with a resolution of $1,000\times 1,000$ pixels. For the calculation of electron temperature, we use both the two-temperature model and the parameterized R-$\beta$ model. The accretion rate is normalized at $230\,\rm GHz$ with an averaged flux of $2.4\,\rm Jy$. We mainly discuss the property of the emission at sub-millimeter frequency, which is dominated by a thermal component. Therefore, in this work, only thermal synchrotron is considered \citep[e.g.,][]{2023arXiv230816740N}.

\begin{figure*}
	\includegraphics[height=.4\linewidth]{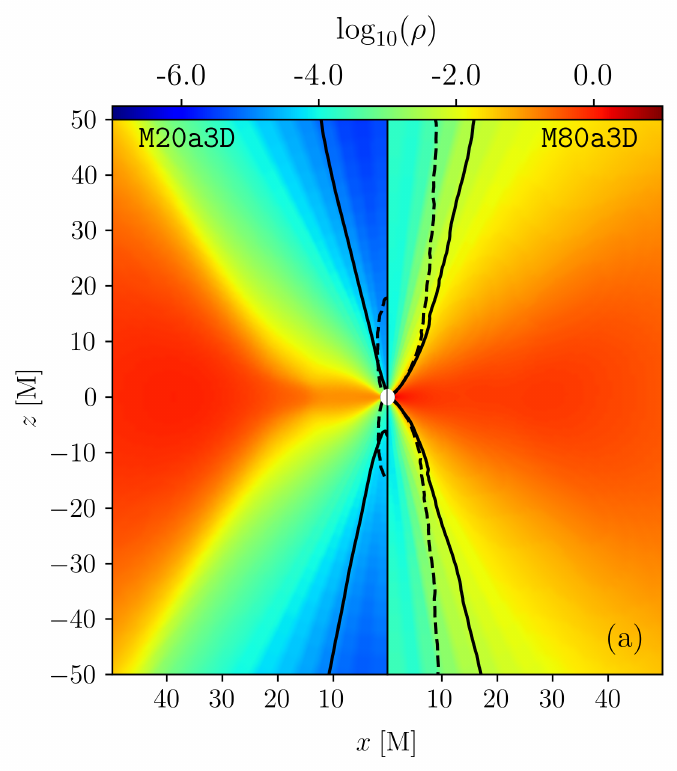}
	\includegraphics[height=.4\linewidth]{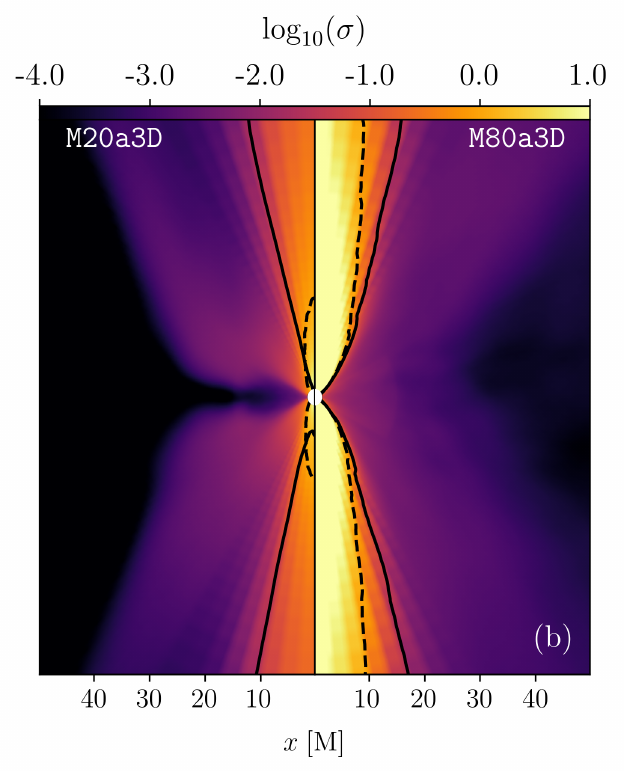}
    \includegraphics[height=.4\linewidth]{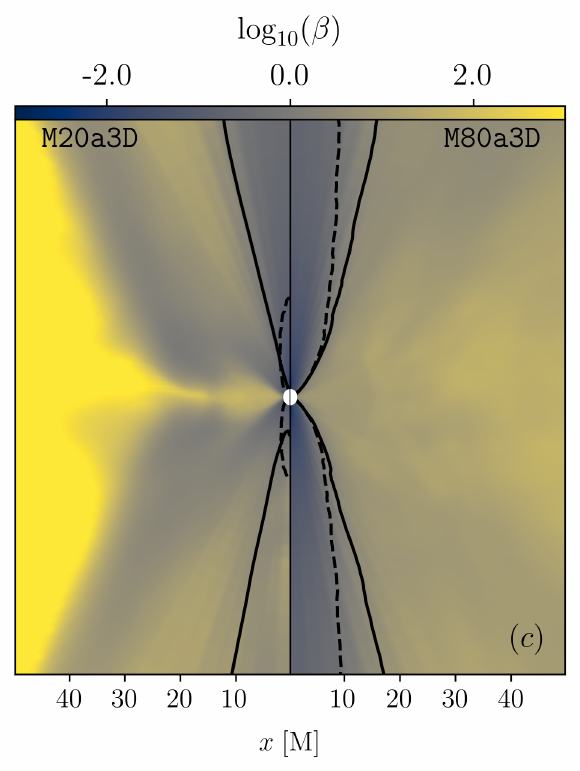}
    \includegraphics[width=.9\linewidth]{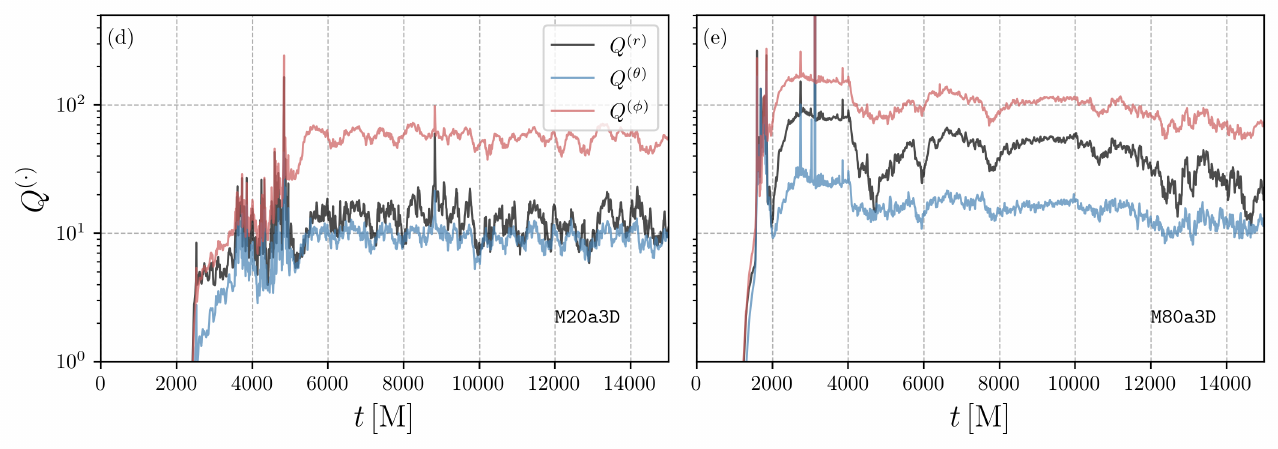}
    \caption{Distribution of time (between $t=10,000\,\rm M$ to $15,000\,\rm M$) and azimuthal averaged (a) density $\rho$, (b) magnetization $\sigma$ and (c) plasma beta for 3D GRMHD simulations with multi-loops in different loop length ({\it left}: $\lambda=20$, {\tt M20a3D} and {\it right}: $\lambda=80$, {\tt M80a3D}). The solid and dashed contours in each panel represent the Bernoulli parameter $-hu_t=1$ and magnetization $\sigma=1$, respectively. Lower panels (d) and (e) show the evolution of averaged MRI quality factors of case {\tt M20a3D} and {\tt M80a3D}.}
    \label{fig: avg_GRMHD}
\end{figure*}
\begin{figure}
    \centering
         \includegraphics[width=\linewidth]{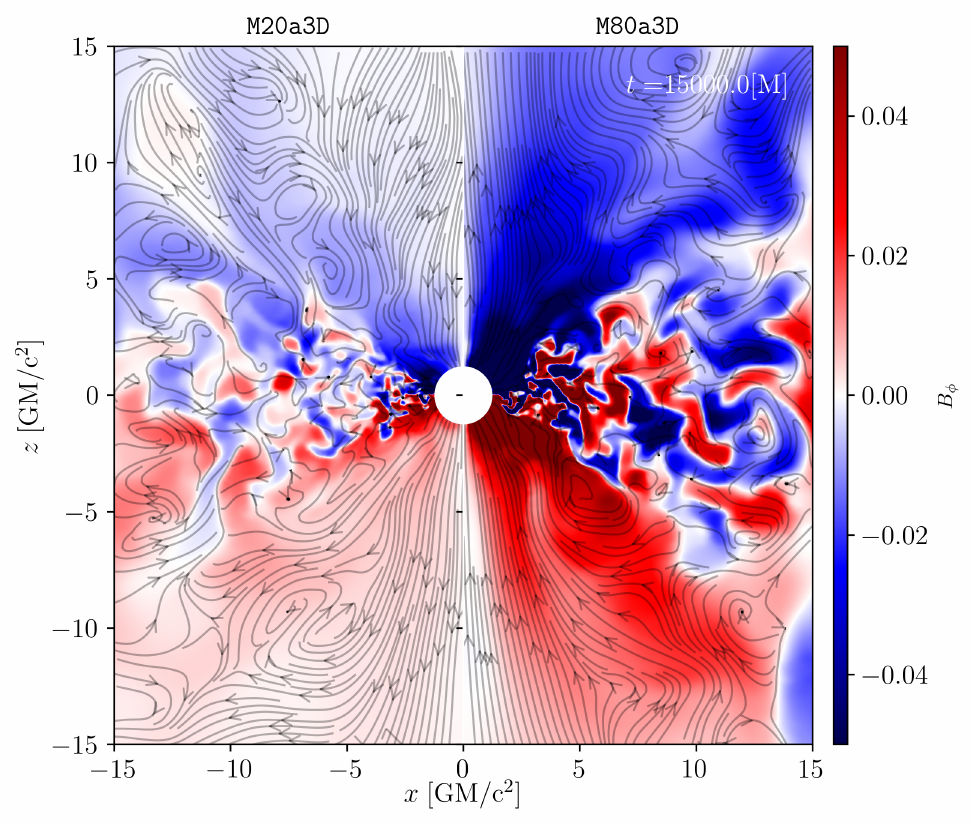}
    \caption{Distributions of toroidal component of the magnetic field for case {\tt M20a3D} ({\it left}) and {\tt M80a3D} ({\it right}). The streamlines show the poloidal magnetic field lines.}
    \label{fig: Bphi}
\end{figure}
\section{Results}
\subsection{Dynamics of accretion flow}

To explore the dynamics of accretion flows in 3D GRMHD simulations of magnetized accretion flows with multiple poloidal loops, we employ the definition from \cite{Porth2019} to compute the mass accretion rate evaluated at the event horizon:
\begin{equation}
    \dot M = \int_0^{2\pi}\int_0^{\pi} \rho u^r \sqrt{-g}d\theta d\phi, \label{Eq: a_rate}
\end{equation}
and the magnetic flux rate measured at the event horizon as,
\begin{equation}
    \Phi_{\rm B} = \frac{1}{2}\int_0^{2\pi}\int_{0}^{\pi}\left|B^r\right|\sqrt{-g}d\theta d\phi. \label{Eq: B-flux}
\end{equation}
Fig.~\ref{fig: Mdot} shows the time evolution of mass accretion rate $\dot M$ and normalized magnetic flux rate $\Phi_{\rm B}/\sqrt{\dot{M}}$ for the cases {\tt M20a3D} (black) and {\tt M80a3D} (blue). 
Due to the slower evolution of the bigger magnetic loop in the case of {\tt M80a3D}, we simulate it up to $20,000\,\rm M$ to study the long-term evolution of accretion dynamics.
The accretion flow of case {\tt M20a3D} loses its initial magnetic configuration and enters a non-linear evolution phase after $5,000\,\rm M$. Thus, we simulate it up to $15,000\,\rm M$ only.

In general, the dynamics of the accretion flow of 3D case {\tt M20a3D} is similar to that in 2D cases shown in \cite{stad1106} (Model {\tt M20a}). When the simulation reaches the quasi-steady phase ($10,000-15,000\,\rm M$), it does not have a persistent jet. A similar phenomenon is observed in the smaller torus simulations in \cite{2022MNRAS.513.5204N}.
In 3D simulations, the accretion flow becomes turbulent in the $\phi$ direction due to the interchange instabilities (see \cite{Begelman2022}), which is absent in 2D simulations. 
The initially ordered magnetic loops are mixed up before accreting into the horizon. The magnetic dissipation due to the multiple loops inside the torus reduces the development of magnetorotational instability (MRI). It weakens the poloidal magnetic field inside the torus, which leads to a lower accretion rate at the event horizon (see the black curve in Fig.~\ref{fig: Mdot}(a)). The average value of the accretion rate for the case {\tt M20a3D} ($\dot M \sim 10^{-1}$) is roughly 2 orders of magnitudes lower than that of the simulations with a single loop as reported by \cite{2021MNRAS.506..741M} ($\dot M \lesssim 10$ see Model $a=0.94$ and $a=0$).
However, when the initial magnetic loop size becomes larger, we observe a magnitude higher accretion rate from case {\tt M80a3D} than {\tt M20a3D}. 
In contrast to the findings from the single loop simulation \citep{2021MNRAS.506..741M}, we observe a significantly reduced magnetic flux in both the {\tt M20a3D} and {\tt M80a3D} cases.
Initial configuration of magnetic loops depends on density structure of torus (see Eq.~\eqref{eq:A_phi}). The strength of each magnetic loop becomes weaker in the larger radius. 
Despite the presence of relatively larger magnetic loops in the case of {\tt M80a3D}, there are at least four magnetic field loops with alternating polarities that hold magnetic energy exceeding $10\%$ of the maximum magnetic energy in the torus. Similarly, for the case {\tt M20a3D}, there are over ten loops with magnetic energy significant enough within the torus. The dynamics of these loops mainly contribute to the overall characteristics of the accretion flow.
More small scale magnetic fields are generated in {\tt M80a3D} as compared to single loop simulation in \cite{2021MNRAS.506..741M}. It leads to roughly half of the magnetic flux in the single loop case and fails to generate a fully MAD accretion flow. However, due to less magnetic dissipation in {\tt M80a3D}, MRI is better resolved, leading to a higher accretion rate than case {\tt M20a3D} which is close to the accretion rate in single loop cases \citep[e.g.,][]{2021MNRAS.506..741M}.

To understand the accretion dynamics, we present the time (between $t=10,000\,\rm M$ to $15,000\,\rm M$) and azimuthally averaged distribution of (a) density ($\log_{10}(\rho)$), (b) magnetization ($\log_{10}(\sigma)$), and (c) plasma beta ($\log_{10}(\beta)$) for both the cases in Fig.~\ref{fig: avg_GRMHD}. The solid and dashed lines in panels (a)-(c) represent the boundaries of the Bernoulli parameter $-hu_t=1$ and magnetization $\sigma=1$, respectively. We also present the time evolution of average MRI quality factors $Q^{(r,\theta,\phi)}$ of {\tt M20a3D} and {\tt M80a3D} cases in Fig.~\ref{fig: avg_GRMHD}(d) and (e). The quality factors are spatially averaged in the high-density torus region, i.e., $r<50\,\rm r_g, 60^\circ<\theta<120^\circ$. In the density distributions of the two cases in Fig.~\ref{fig: avg_GRMHD}(a) case {\tt M20a3D} has a larger gravitationally bounded region (see the solid lines) and smaller disk region as compared to case {\tt M80a3D}. The initial magnetic configuration with larger loops produces a less chaotic magnetic field and magnetic dissipation by reconnection inside the torus \citep{stad1106}. Therefore, we observe a stronger jet and lower plasma $\beta$ (stronger magnetic field) in the case {\tt M80a3D} than that of the case {\tt M20a3D} (see Fig.~\ref{fig: avg_GRMHD}(b) and (c)). The different strengths of the leftover magnetic field after the merger of the magnetic loops generate different magnitudes of MRI. In panels Fig.~\ref{fig: avg_GRMHD}(d) and (e), we see that all MRI quality factors are higher than $8$, which means MRI is fully resolved \citep{2011ApJ...738...84H, Porth2019}. However, slightly lower MRI quality factors are observed in case {\tt M20a3D} than case {\tt M80a3D}, which is caused by the globally weaker magnetic field in the torus. This also results in different accretion rates in both the models (Fig.~\ref{fig: Mdot}), and due to that, the case {\tt M20a3D} forms a narrower high-density equatorial region compared to the case {\tt M80a3D}.

In Fig.~\ref{fig: Bphi}, we show the distribution of the toroidal component of the magnetic field ($B_\phi$) along with the poloidal magnetic field lines. In both cases, we observe the formation of opposite polarity islands (indicated by red regions inside the blue region and vice versa). These are the results of active MRI in the accretion flow. We see a stronger toroidal component of the magnetic field close to the horizon for the case {\tt M80a3D} than that of the case {\tt M20a3D}.
It helps for the formation of a stronger jet in the case {\tt M80a3D} as compared to the case {\tt M20a3D} (see Fig.~\ref{fig: avg_GRMHD}(b)).

\subsection{Radiative property}
In this section, we study the radiative signatures of the multi-loop magnetic field configuration in the accretion flows. Earlier, radiative signatures of single-loop magnetic field configuration in the accretion flows have been studied extensively in the two-temperature and in R-$\beta$ considerations for electron temperature \citep[e.g.,][]{Moscibrodzka2016, 2021MNRAS.506..741M, Fromm2022}. More frequent reconnection in multi-loop cases generates more turbulent accretion flow and intermittent strength of jets. In this section, we investigate these radiative properties in detail and study their observational implications.

In this paper, we only consider thermal synchrotron emission to investigate the radiative property of the accretion flow. We calculate emissivity following \cite{2011ApJ...737...21L}, which introduces an approximate expression of the thermal synchrotron emission. The explicit expression can be written as follows:
\begin{equation}
    j_{\rm \nu}=n_{\rm e} \frac{\sqrt{2}\pi e^2 \nu_{\rm s}}{2K_{\rm 2}\left(1/\Theta_{\rm e}\right)c}\left(X^{1/2}+2^{11/12}X^{1/6}\right)^2\exp{\left(-X^{1/3}\right)}, \label{Eq: MBS_emissivity}
\end{equation}
where 
\begin{equation}
    X\equiv \frac{\nu}{\nu_{\rm s}},\, \nu_{\rm s} \equiv (2/9)\nu_{\rm c}\Theta_{\rm e}^2\sin{\theta} \label{Eq: X}.
\end{equation}
In the above equation, $\Theta_{\rm e}=k_{\rm B} T_{\rm e}/m_{\rm e} c^2$ is the dimensionless electron temperature, $k_{\rm B}$ is the Boltzmann constant, $T_{\rm e}$ is the electron temperature in CGS unit, $K_2$ is the modified Bessel function of the second kind, and $\nu_{\rm c}$ is the electron cyclotron frequency, which is given by,
\begin{equation}
    \nu_{\rm c} \equiv \frac{eB}{2\pi m_{\rm e} c}=2.8\times 10^6B\,\rm{Hz}, \label{Eq: nu_c}
\end{equation}
where the symbols have their usual meaning, viz., $e$ is the electronic charge, $m_{\rm e}$ is the electronic mass, magnetic field strength $B=\sqrt{B^i B_i}$, and $B^i$ is the Eulerian 3-magnetic field. 

In our calculations, we use R-$\beta$ prescription as well as two-temperature models to calculate the electron temperature in the accretion flow. In the R-$\beta$ prescription, the electron temperature is calculated from the proton temperature as follows \citep{Moscibrodzka2016}:
\begin{equation}
    \frac{T_{\rm p}}{T_{\rm e}} = \frac{1}{1+\beta^2}R_{\rm l}+\frac{\beta^2}{1+\beta^2}R_{\rm h},
    \label{Eq: R-beta}
\end{equation}
where $R_{\rm l}$ and $R_{\rm h}$ are two dimensionless parameters. In our work, we fix the $R_{\rm l}=1$ and use two different values for $R_{\rm h}=1$ and $160$ to compare with the two-temperature electron heating models.

\begin{figure}
    \centering
         \includegraphics[width=\linewidth]{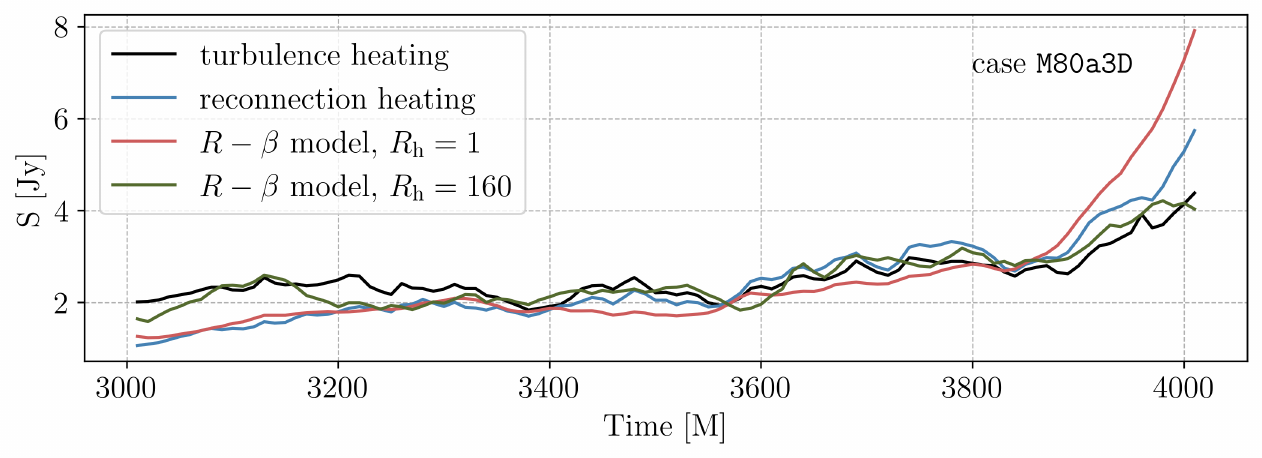}
    \caption{Evolution of light curves at $230\,\rm{GHz}$ for the case of $\tt M80a3D$ with different heating models and R-$\beta$ models. The black solid line represents the light curve from the turbulence heating model. The blue solid line is the reconnection heating one. The red and dark-green lines represent the R-$\beta$ models with $R_{\rm h}$ equals 1 and 160, respectively.}
    \label{fig: case6_lc_300-400}
\end{figure}
\begin{figure}
	\includegraphics[width=\linewidth]{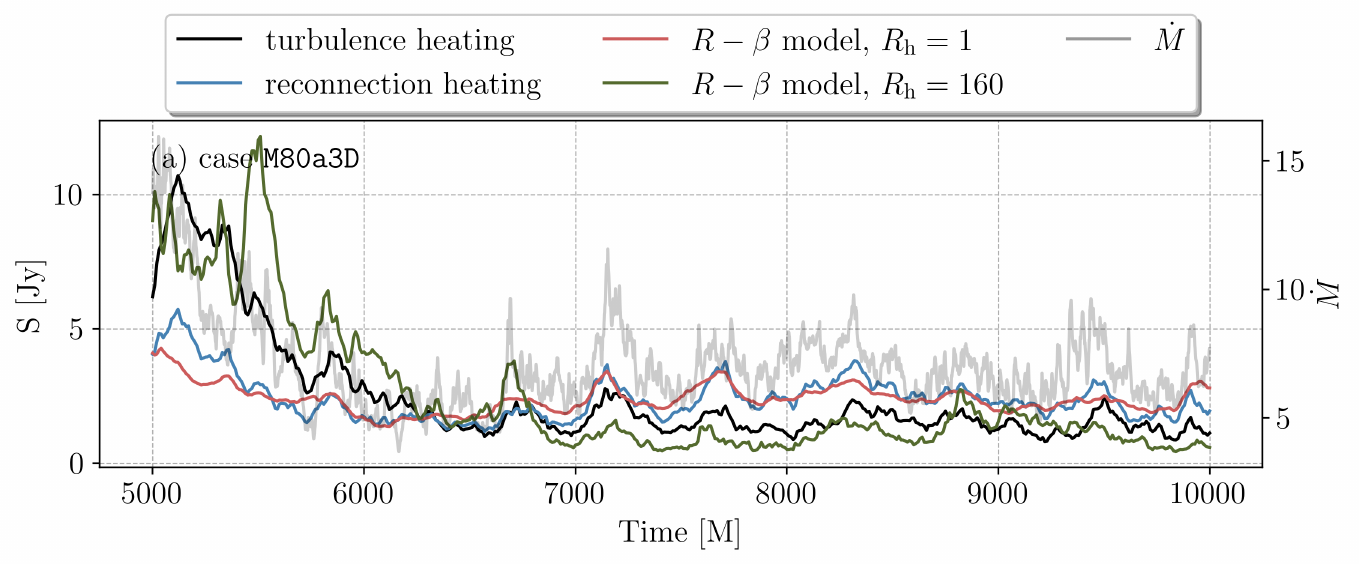}
	\includegraphics[width=.99\linewidth]{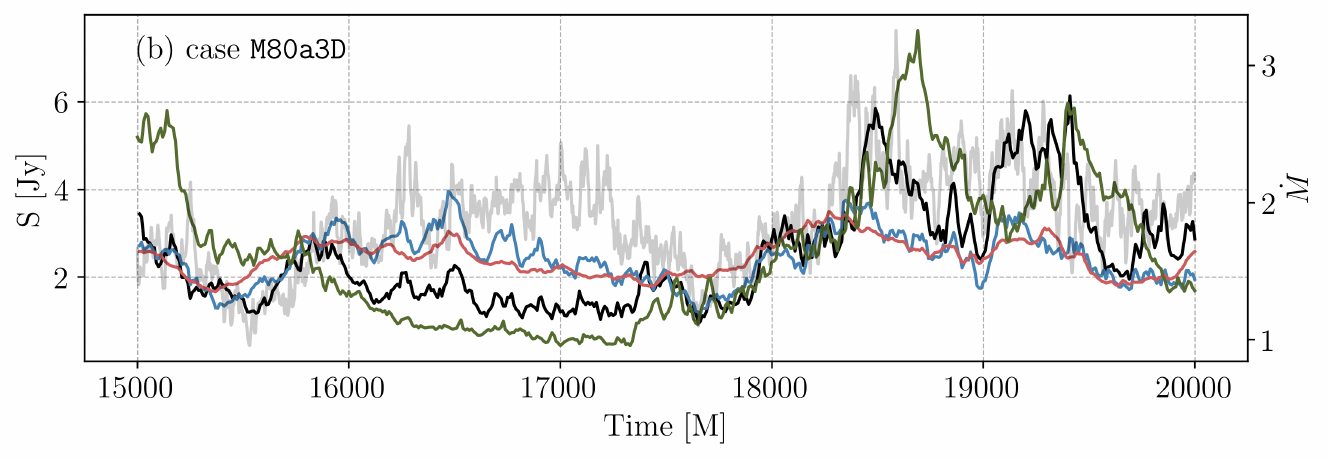}
	\includegraphics[width=\linewidth]{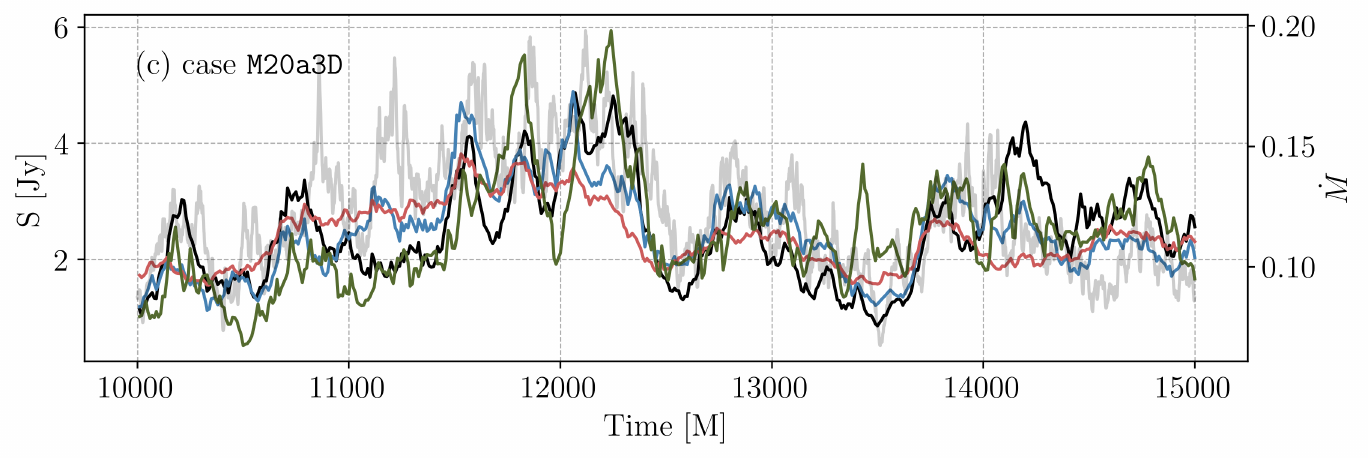}
    \caption{Panel (a) and (b) are the light curves of 4 different electron heating models for the case $\tt M80a3D$, which is the same as Fig.~\ref{fig: case6_lc_300-400} but in different times. Panel (c) is for the case $\tt M20a3D$. In each panel, the grey solid line is the corresponding accretion rate during each period.}
    \label{fig: flare_lc}
\end{figure}
\begin{figure*}
\centering
	\includegraphics[width=.9\linewidth]{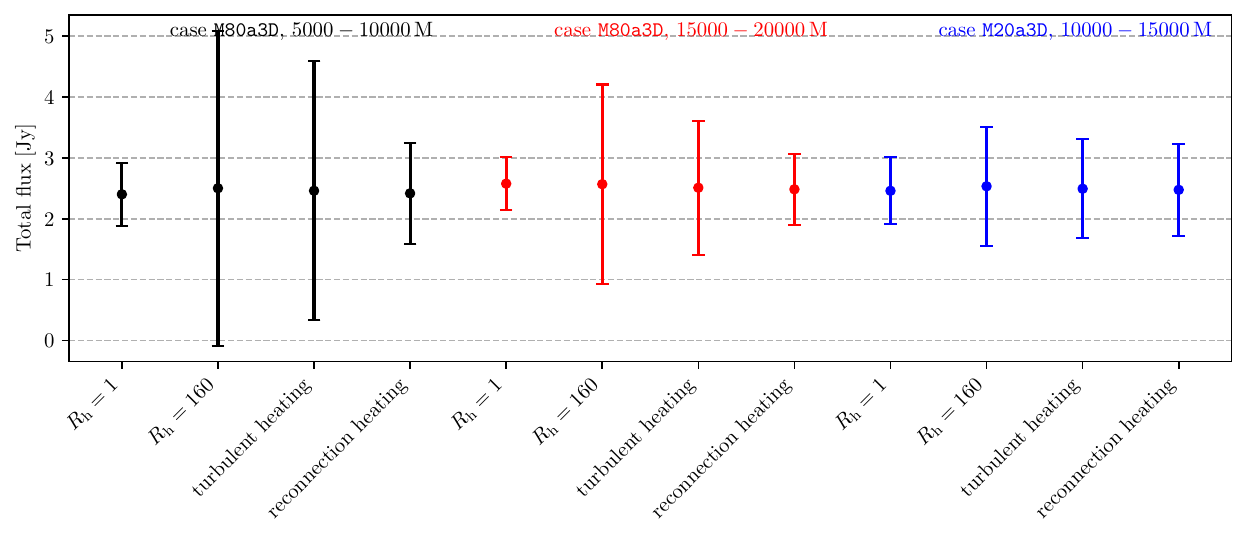}
    \caption{Standard deviation of the light curves of different electron heating models. The error bar represents the standard deviation of the light curves. Black dots show the deviation of case $\tt M80a3D$ during $5,000-10,000\,\rm M$, red dots are $15,000-20,000\,\rm M$, and blue dots show the data of case $\tt M20a3D$ during $10,000-15,000\,\rm M$.}
    \label{fig: std}
\end{figure*}
%

\subsubsection{Light curves}
In Fig.~\ref{fig: case6_lc_300-400}, we show the light curves of the case $\tt M80a3D$ at 230 GHz during the initial phase of the evolution. The black, blue, red, and green lines correspond to the light curves calculated for turbulence heating, reconnection heating, and the R-$\beta$ model with $R_{\rm h}=1$ and $R_{\rm h}=160$, respectively.
During the early phase of the evolution, accretion happens only due to the first magnetic loop inside the torus (simulation time $t=3,000-4,000\,\rm M$), which is quite similar to the single-loop simulations.
During this phase, the light curves of different electron heating models and R-$\beta$ models show similar behavior with small variations. That is because the reconnection between the poloidal field lines with different polarities starts only after $t\gtrsim4,000\,\rm M$. This result agrees with \cite{2021MNRAS.506..741M} due to the similarity of the magnetic field configuration. 
However, when the follow-up loops with different polarities accrete onto the horizon, the behavior of the light curves from different electron heating models varies. During the period that is $t\gtrsim 4,000\,\rm{M}$, we see strong magnetic reconnection between the first and second loops, which releases a large amount of magnetic energy and contributes to increasing the luminosity. 

Fig.~\ref{fig: flare_lc} shows the light curves in different electron heating models at the later simulation time for both the cases ({\tt M80a3D}, {\tt M20a3D}). The behavior of the light curves from different electron temperature models shows significant differences from the ones at the earlier simulation time shown in Fig.~\ref{fig: case6_lc_300-400}. In Fig.~\ref{fig: flare_lc}, we see that the global trend of the light curve for $\rm R-\beta$ models and the two-temperature models is similar, which roughly follows the trend of mass accretion rate. 
However, the detailed structure of the light curves for each electron heating model is different. The turbulence heating model is closer to the result of the $R_{\rm h} = 160$ case, while reconnection heating is closer to the $R_{\rm h} = 1$ case. The two different values of $R_{\rm h}$ of the R-$\beta$ model represent two extreme conditions of the electron heating prescriptions. The effect of different $R_{\rm h}$ values is mainly on the estimated electron temperature of the high plasma $\beta$ region. A higher $R_{\rm h}$ value gives a lower temperature in the disk. 
Therefore, in that case, we observe less radiation from the disk as compared to the lower $R_{\rm h}$ value case. 
%
If the turbulence and frequent reconnection do not happen, the difference of the 230 GHz light curves between the two-temperature and R-$\beta$ models is expected to be minimal, as seen in the initial simulation time (Fig.~\ref{fig: case6_lc_300-400}) and previous single loop simulations \citep{2021MNRAS.506..741M}.

To understand the variability of the 4 different models for both cases, in Fig.~\ref{fig: std}, we show the standard deviations of light curves at 230 GHz for different electron heating models and R-$\beta$ models at different time periods. The first two sections of Fig.~\ref{fig: std} correspond to the case {\tt M80a3D}, but at different periods of time, $t=5,000-10,000\,$M and $t=15,000-20,000\,$M. The rightmost section of Fig.~\ref{fig: std} corresponds to the case {\tt M20a3D} in the period of $t=10,000-15,000\,$M.
In general, the R-$\beta$ model with $R_{\rm h}=160$ shows the highest standard deviation, suggesting it is the most variable case among the 4 models. On the contrary, the light curve for $R_{\rm h}=1$ is the least variable model. By comparing panels (a) and (b) in Fig.~\ref{fig: flare_lc}, we observe that due to the continuous merging of the magnetic field loops, the variability reduces with simulation time. It is reflected in Fig.~\ref{fig: std} that all models have lower standard deviations at a later simulation time. Similarly, by comparing panels (c) with (a) in Fig.~\ref{fig: flare_lc}, we find that for case $\tt M80a3D$, due to the larger and stronger magnetic field loops in the initial torus, the variability of light curves is larger than that of case $\tt M20a3D$. In Fig.~\ref{fig: std}, case $\tt M20a3D$ has a smaller averaged standard deviation than that of the case $\tt M80a3D$.

\begin{figure}
        \centering
	\includegraphics[width=\linewidth]{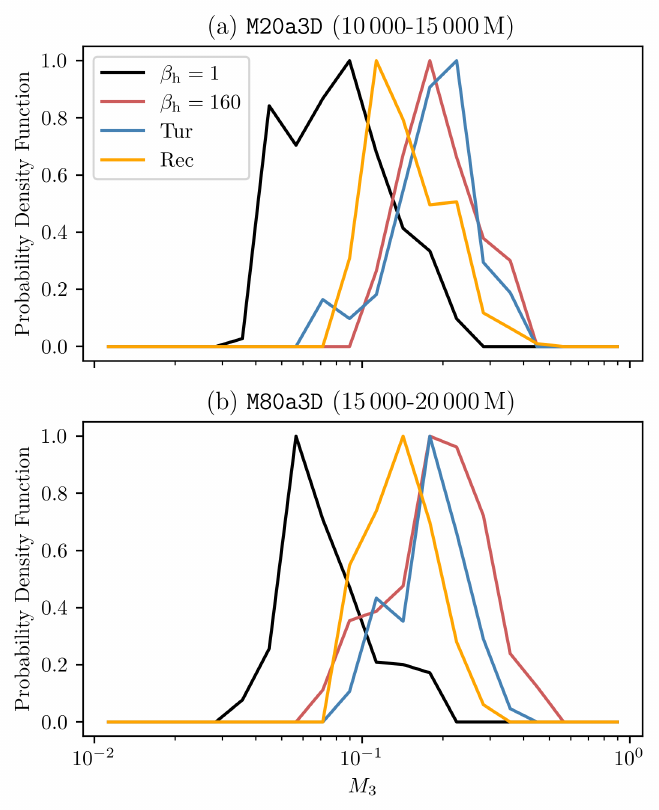}
    \caption{Probability distribution function (PDF) of the $M_3$ of the light curves generated from case $\tt M20a3D$ (a) and $\tt M80a3D$ (b). The PDFs are normalized by setting the peak value to be 1. The black and red lines represent the cases with different $R_{\rm h}$ values from the $R-\beta$ model. The blue and yellow lines are from turbulence and reconnection models, respectively.}
    \label{fig: PDF_M3}
\end{figure}

Other than the standard deviation in the light curves, in \citet{event_horizon_telescope_collaboration_first_2022}, a modulation index $M_3$ with a duration of $3\,\rm hr$ was utilized to quantify the light curve variation. The modulation index $M_3$ is defined as the ratio between the standard deviation $\sigma_{3\,\rm hr}$ and the mean value $\mu_{3\,\rm hr}$ calculated over a time bin lasting approximately $3\,\rm hr$ from the light curve:
\begin{equation}
    M_3\equiv \frac{\sigma_{3\,\rm hr}}{\mu_{3\,\rm hr}}
\end{equation}
It offers quantitative measurement of the light curve variability. Historical $M_3$ observation suggests $M_3\lesssim0.1$. However, most of the simulation models in \cite{event_horizon_telescope_collaboration_first_2022} failed to match this constraint. The $M_3$ distribution from the light curves of 4 different electron heating models of the cases {\tt M80a3D} and {\tt M20a3D} is presented in Fig.~\ref{fig: PDF_M3}. Similar to Fig.~\ref{fig: std}, the turbulence model shows a similar distribution as the $R-\beta$ model with $R_{\rm h}=160$. The $R-\beta$ model with $R_{\rm h}=1$ and the reconnection heating model have lower variability. In particular, the $R-\beta$ model with $R_{\rm h}=1$ presents a similar value ($\sim 0.06$) of historical observation of Sgr\,A$^{*}$.

The difference among the light curves from different electron heating models suggests that for different models, the dominant emitting region is different. It is determined by the electron temperature distribution. We will discuss it in detail in the subsequent sections.

\subsection{GRRT images at the flaring state}
\subsubsection{Emitting region at 230\,GHz}

\begin{figure*}
\centering
        \includegraphics[height=.275\linewidth]
        {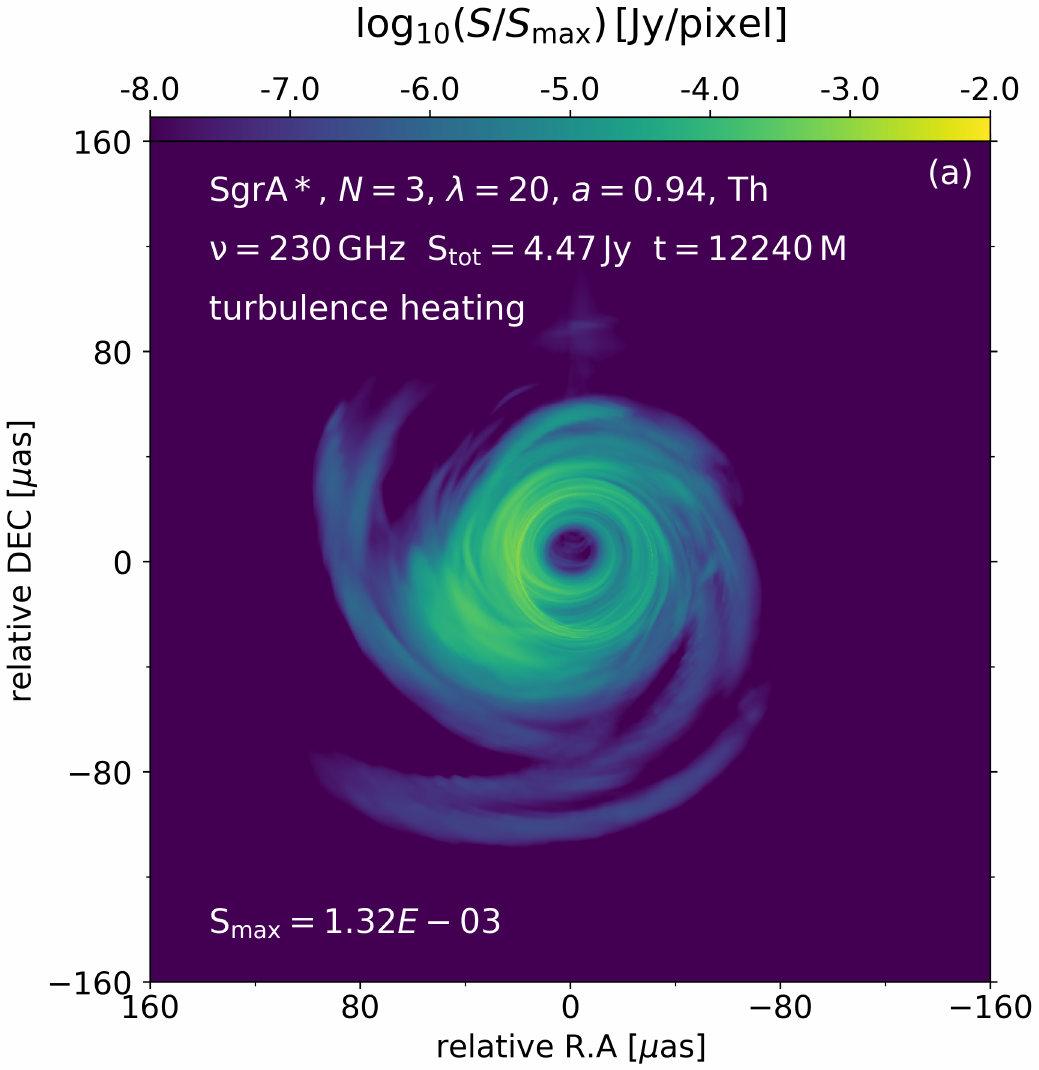}
	\includegraphics[height=.275\linewidth]{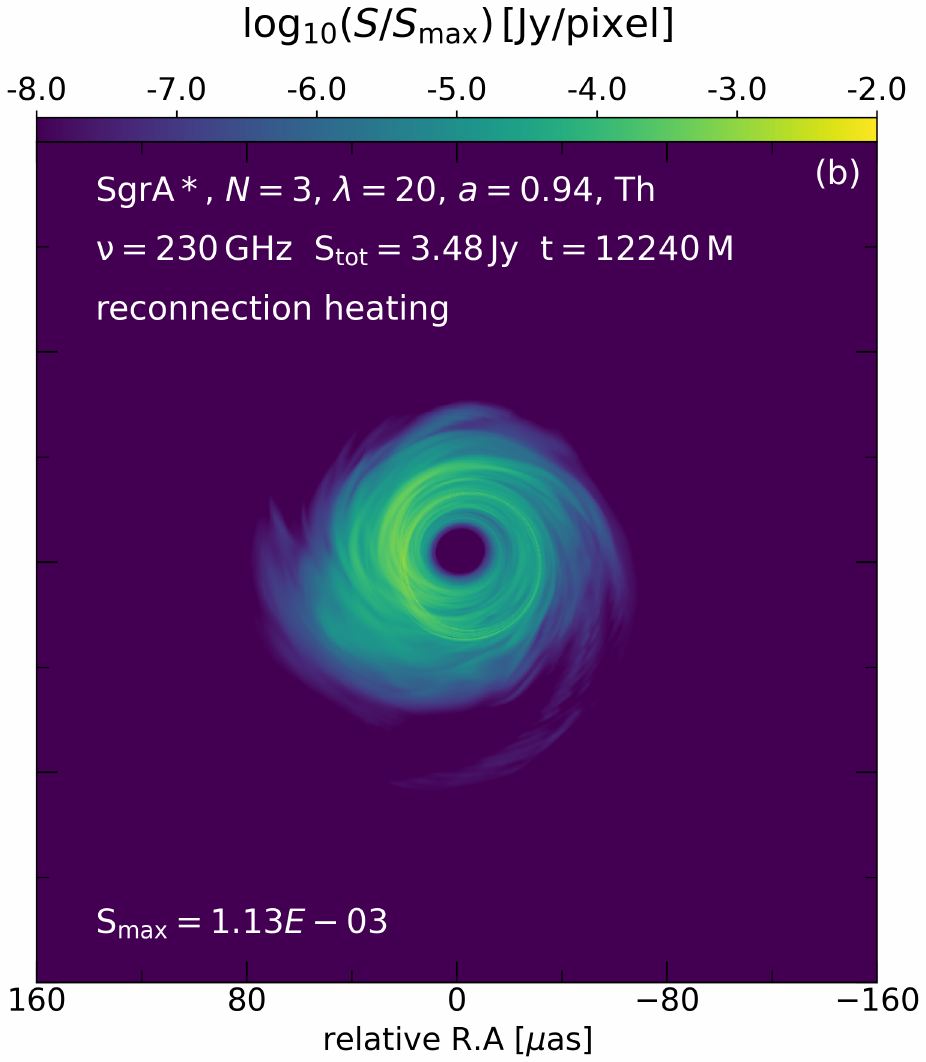}
	\includegraphics[height=.275\linewidth]{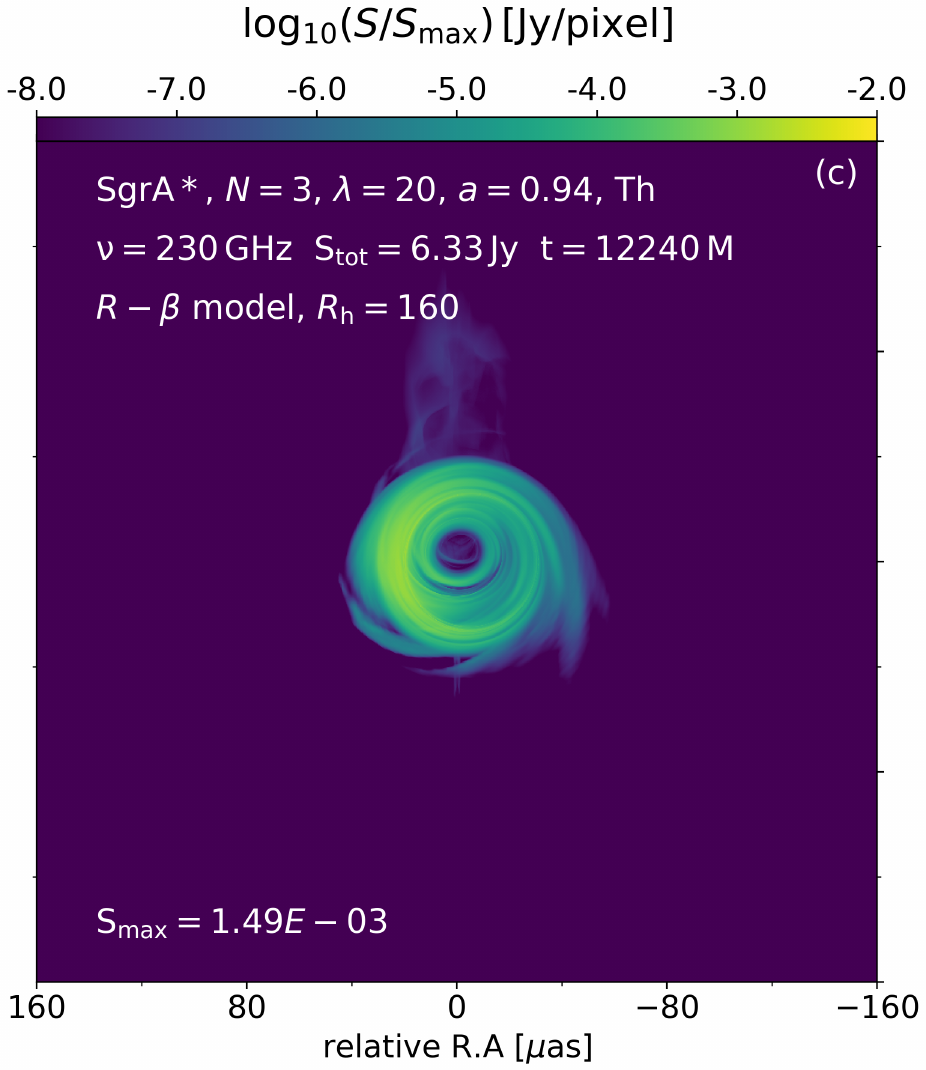}
	\includegraphics[height=.275\linewidth]{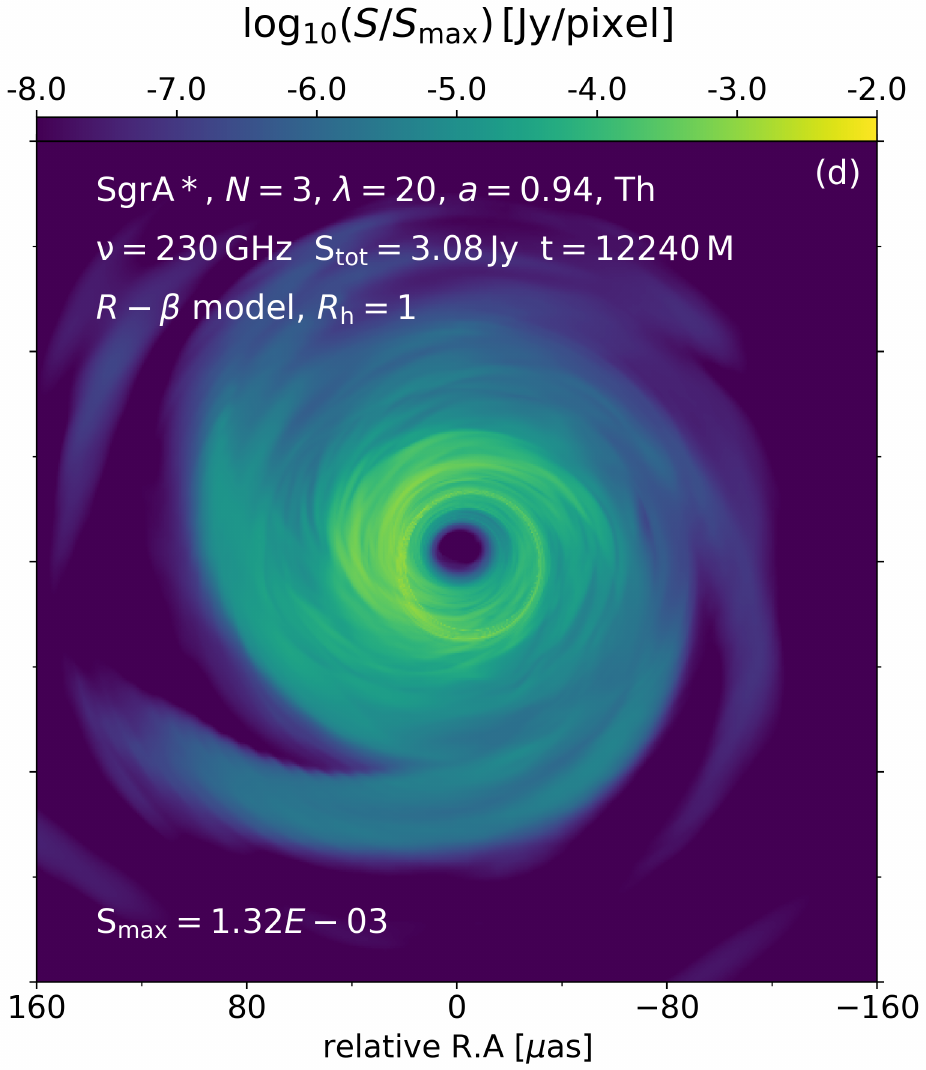} 
 \centering
	\includegraphics[height=.31\linewidth]{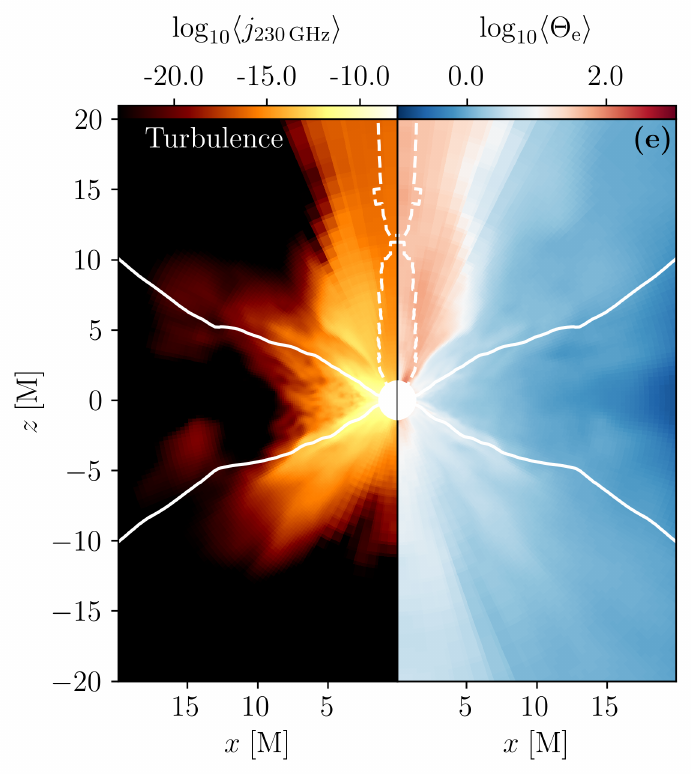}
	\includegraphics[height=.31\linewidth]{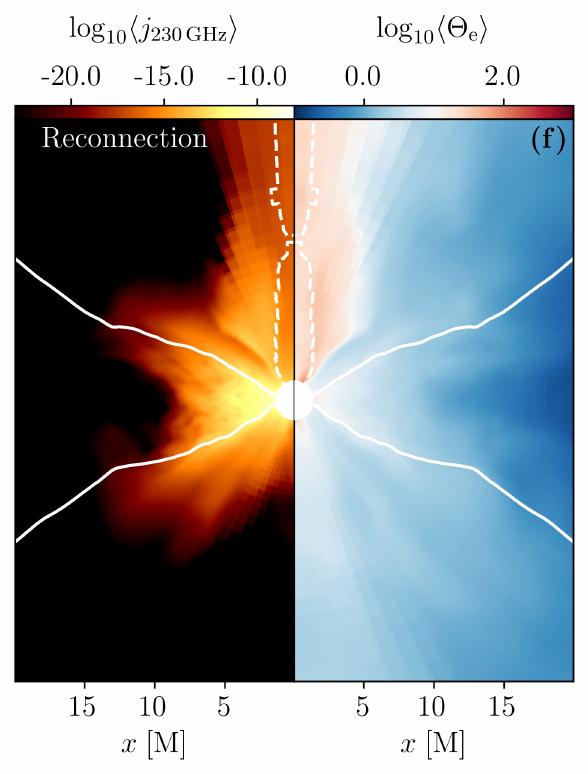}
	\includegraphics[height=.31\linewidth]{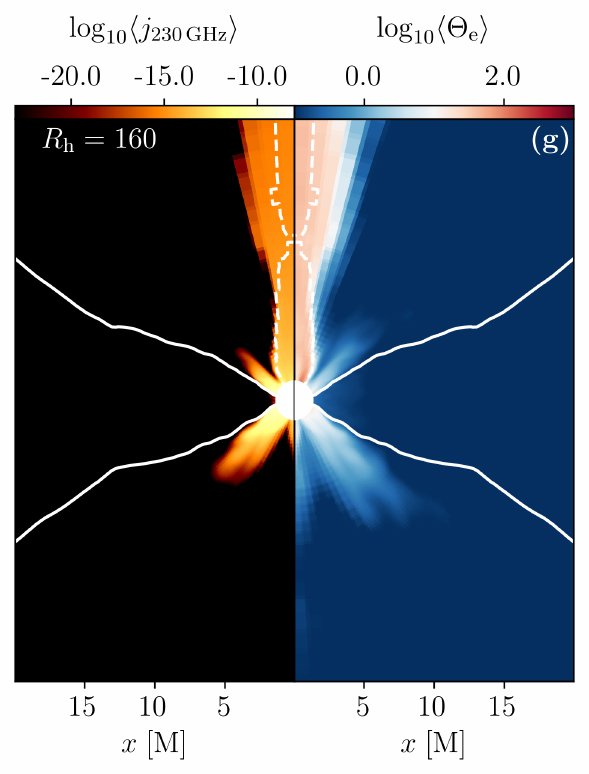}
	\includegraphics[height=.31\linewidth]{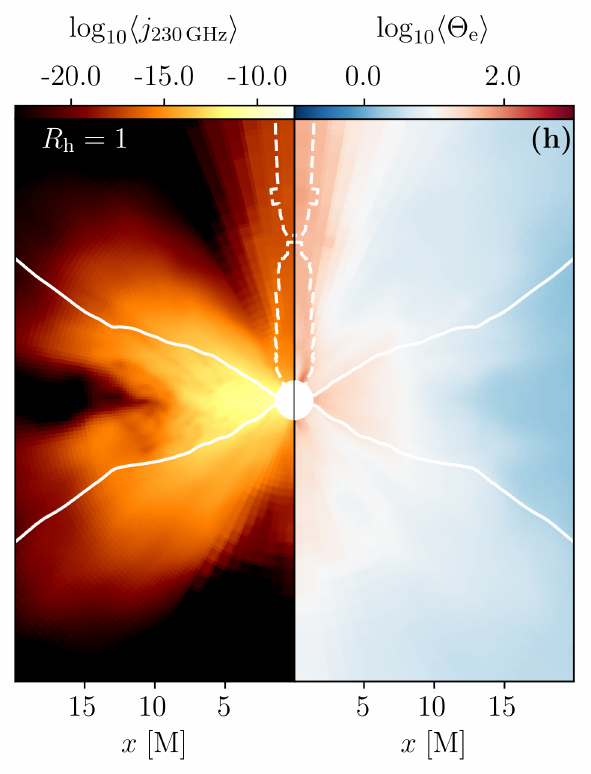}
	
    \caption{The panels in the first row are the GRRT snapshot images for Sgr\,A$^*$ with an inclination angle $i=30^\circ$ at $t=12,240\,\rm M$ from (a) turbulence heating model, (b) reconnection heating model, (c) $\rm R-\beta$ model with $R_{\rm h} = 160$ and panel (d) $\rm R-\beta$ model with $R_{\rm h}=1$. The second row is the corresponding $230\,\rm GHz$ emissivity ({\it left}) and electron temperature ({\it right}) distributions. The solid white contours in panel (e-h) represent the edge of the torus which is the $\rho=0.01$ contour. The dashed contour represents the high magnetization region with $\sigma>1$, which is cut off in GRRT post-process. 
    }
    \label{fig: Thetae}
\end{figure*}

In the previous section, we demonstrate the distinct features of light curves for different electron heating models, which could be essential to understanding the variability observed in Sgr\,A$^*$ \citep{event_horizon_telescope_collaboration_first_2022-1}.
The dynamics and the dominant emitting region of the accretion flow are related to the variability of the observed light curves. In this section, we investigate the connection between the GRRT images from different electron heating models and the corresponding emissivity distributions in the GRMHD simulations.

As shown in Fig.~\ref{fig: Bphi}, the case {\tt M20a3D} generates more turbulence in the accretion flows and occurs more reconnections than those in the case {\tt M80a3D}, which leads to the eruption of more flux ropes \citep{2022ApJ...933...55C}. We expect to relate it to the flaring events as seen in Sgr\,A$^*$. To study the properties of the flux ropes and their connection to the observed flaring events, we focus our discussion on case $\tt M20a3D$ in the following sections.

In the first row of Fig~\ref{fig: Thetae}, we show the snapshot GRRT images of 230 GHz for the 4 different electron heating models of the case {\tt M20a3D} at $t=12,240\, \rm M$. 230 GHz snapshot images from two different electron heating models (turbulence and reconnection) are presented in panels (a) and (b), respectively. 
The turbulence model creates a more extended filamentary structure in the GRRT image than the reconnection model. We see snapshot GRRT images from the R-$\beta$ model with $R_{\rm h}=160$ and $1$ in panels (c) and (d), respectively. The $R_{\rm h}=160$ case shows a very compact GRRT image, while the $R_{\rm h}=1$ one has the most extended emission among the 4 cases.

The differences in the GRRT images seen in Fig.~\ref{fig: Thetae} are due to the different electron temperature distributions of the models. To understand it, in the second row of Fig.~\ref{fig: Thetae}, we present the corresponding distributions of azimuthally averaged emissivity ({\it left}) and electron temperature ({\it right}) for different electron heating models of GRRT images. 
In general, the two-temperature models (turbulence and magnetic reconnection heating) agree with each other qualitatively in both distributions of electron temperature and emissivity. The difference comes from the more extended emission in the sheath region from the turbulence model, which is the source of the more extended filament structure in the GRRT image. The emission for the reconnection heating model is more concentrated from the disk near the equatorial plane.

The electron temperature distributions of the R-$\beta$ model vary depending on the values of $R_{\rm h}$. For the case, $R_{\rm h}=160$, the contribution from the disk is very low (see Fig.~\ref{fig: Thetae}(g)). Therefore, similar to the turbulence heating model, the emission at $230\,\rm GHz$ mainly comes from the sheath region. 
The case of $R_{\rm h}=1$ provides more extended emission due to higher disk temperature (see Fig.~\ref{fig: Thetae}(h)), which corresponds to the large and extended diffuse emission in the GRRT image seen in Fig~\ref{fig: Thetae}(d). A further detailed comparison of electron temperature distribution for different electron heating models is discussed in Appendix~A.

\begin{figure*}
    \centering
	\includegraphics[width=.8\linewidth]{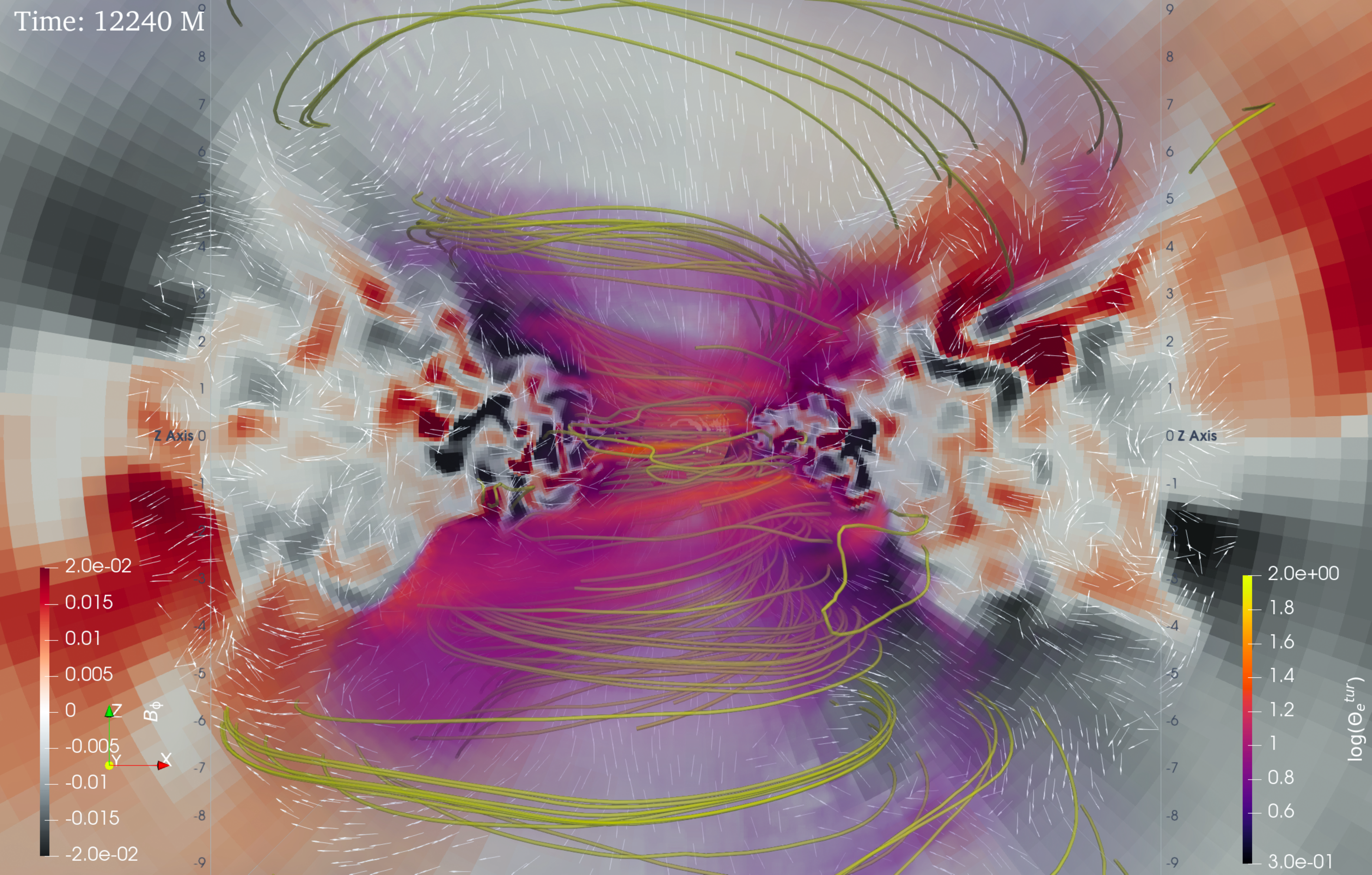}
    \caption{Volume rendering of 3D distribution of electron temperature (yellow to black) and toroidal component of the magnetic field (red to black) of the turbulence heating model at $12,240\,\rm M$ of case {\tt M20a3D}. The white arrows mark the direction of the poloidal magnetic field in it. Yellow tubes represent the magnetic lines. The size of the box is $10\,\rm r_{g}$.}
    \label{fig: flare_Thetae}
\end{figure*}

The emission from the jet sheath is interesting to study as most of the plasmoid chain \citep{stad1106,nathanail_plasmoid_2020} and flux ropes \citep{2022ApJ...933...55C} are generated there. A similar phenomenon was also observed in GRPIC simulation in \cite{Mellah2023}. In general, the flux ropes contain a relatively more ordered magnetic field as well as higher density and electron temperature. In Fig.~\ref{fig: flare_Thetae}, we show the 3D distribution of electron temperature from the turbulence heating model for the case {\tt M20a3D} at $t=12,240\,\rm M$. 
The volume rendering of electron temperature is seen in orange and purple colors. It clearly shows the filamentary structure (red and purple) in the sheath region. It is a flux rope and contains ordered bundles of magnetic field lines (indicated in yellow magnetic lines). The red and black background in Fig.~\ref{fig: flare_Thetae} shows the distribution of the toroidal component magnetic field. Flux ropes emerge near the horizon within the jet sheath region through magnetic reconnection, eventually forming a helical structure. We consider a similar mechanism for the formation of the plasmoid chain as seen in 2D high-resolution simulation (e.g., \cite{stad1106}). From Fig.~\ref{fig: flare_Thetae} we see a relatively strong magnetic field at the edge of a torus with different polarities, which causes strong reconnection. Due to the low numerical resolution in our 3D simulations than that in 2D simulations, tearing instability is not able to develop. Therefore, we do not see clear evidence of the development of plasmoids in our 3D simulations. 
However, if we perform the 3D simulations with very high numerical resolution, plasmoid chains can be seen \citep{Ripperda2021}. Unfortunately, due to the limitations of our numerical resources, we can not perform such super-high-resolution simulations. We think a reasonable conjecture is that the flux ropes observed in Fig.~\ref{fig: flare_Thetae} would be unresolved plasmoid chains.

\begin{figure}
    \centering
	\includegraphics[width=\linewidth]{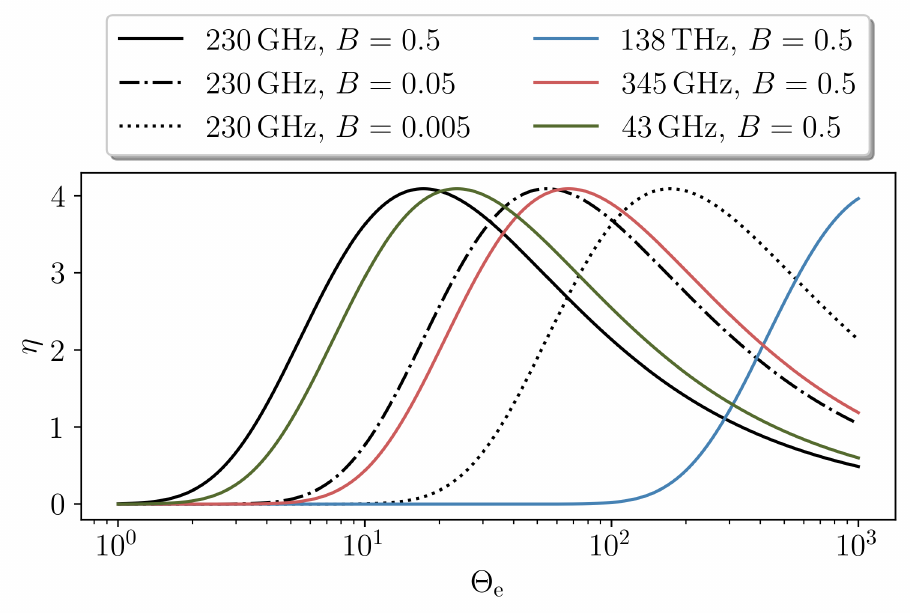}
    \caption{Distribution of emissivity coefficient $\eta$ as a function of electron temperature. Solid lines represent the frequency of $230\,\rm GHz$ with different local magnetic field strengths. The purple dash-doted, red dashed, and dotted lines represent the emissivity coefficient by 46 GHz, 86 GHz, and NIR frequencies, respectively, with typical magnetic fields in flux rope ($B=0.05$). 
    }
    \label{fig: emissivity}
\end{figure}

\subsubsection{Filamentary structure and flux ropes}

In the last section, we see filamentary structures in GRRT images obtained from the turbulence heating model. This emission originates from emerging flux ropes. To establish this correlation rigorously, we deeply investigate the interplay between emissivity and electron temperature. According to Eq.~\ref{Eq: MBS_emissivity}, thermal synchrotron emission $j_{\rm \nu}$ is proportional to electron number density $n_{\rm e}$, which is related to density $\rho$ in the simulations. From Fig.~\ref{fig: Thetae}, the funnel region near the pole has a high electron temperature. However, due to the low density, the contribution to the emission from the funnel region is small.

Under ultra-relativistic condition ($\Theta_{\rm e}\gg1$), we have \citep{2011ApJ...737...21L}:
\begin{equation}
    K_{\rm 2}\left(1/\Theta_{\rm e}\right)\sim \Theta_{\rm e}^2.
\end{equation}
Different electron heating models give different distributions of electron temperature. Thus, we focus on the dependency of electron temperature on the emissivity. Leaving other parameters such as electron number density, $n_{\rm e}$ and magnetic field strength, $B$ as constant, we rewrite Eq.~\ref{Eq: MBS_emissivity} as follows:
\begin{equation}
    j_{\rm \nu}\propto \eta(\Theta_{\rm e}, B)=\left(X^{1/2}+2^{11/12}X^{1/6}\right)^2\exp{\left(-X^{1/3}\right)}, \label{Eq: gamma}
\end{equation}
where $X\equiv \nu/\nu_{\rm s}\sim \Theta_{\rm e}^{-2}$. 
Here we define $\eta$ as a coefficient proportional to emissivity. From Eq.~\ref{Eq: X} and \ref{Eq: nu_c}, the relation between emissivity $j_{\rm \nu}$ and electron temperature $\Theta_{\rm e}$ is only influenced by observing frequency $\nu$ and the local magnetic field $B$ (see also \cite{event_horizon_telescope_collaboration_first_2022}). Figure \ref{fig: emissivity} shows the emissivity coefficient as a function of electron temperature in different magnetic field strengths and frequencies. This helps us understand the relationship between emissivity and electron temperature. It indicates that the strength of the local magnetic field is very sensitive to emissivity.

%
\begin{figure}
    \centering
	\includegraphics[width=\linewidth]{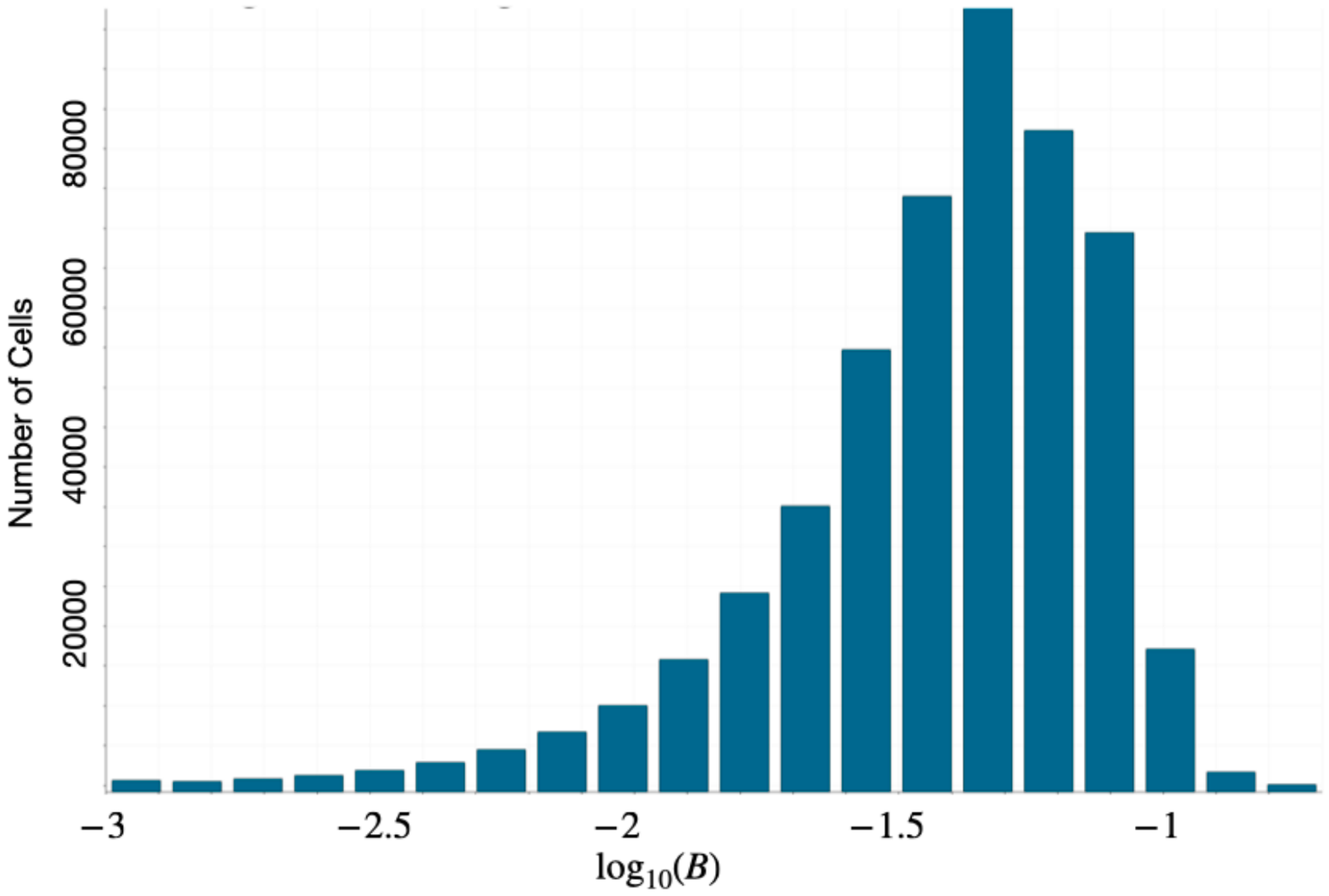}
    \caption{Histogram of magnetic field strength distribution in the flaring region. Here, the unit of magnetic field is in code unit. We measure the magnetic field strength at high electron temperature ($\Theta_{\rm e}>10$) and high density ($\rho>0.001$) region at $t=12,240\,\rm M$ of the turbulence heating model of case {\tt M20a3D}.} 
    \label{fig: B_distribution}
\end{figure}

\begin{figure}
    \centering
    \includegraphics[width=.8\linewidth]
{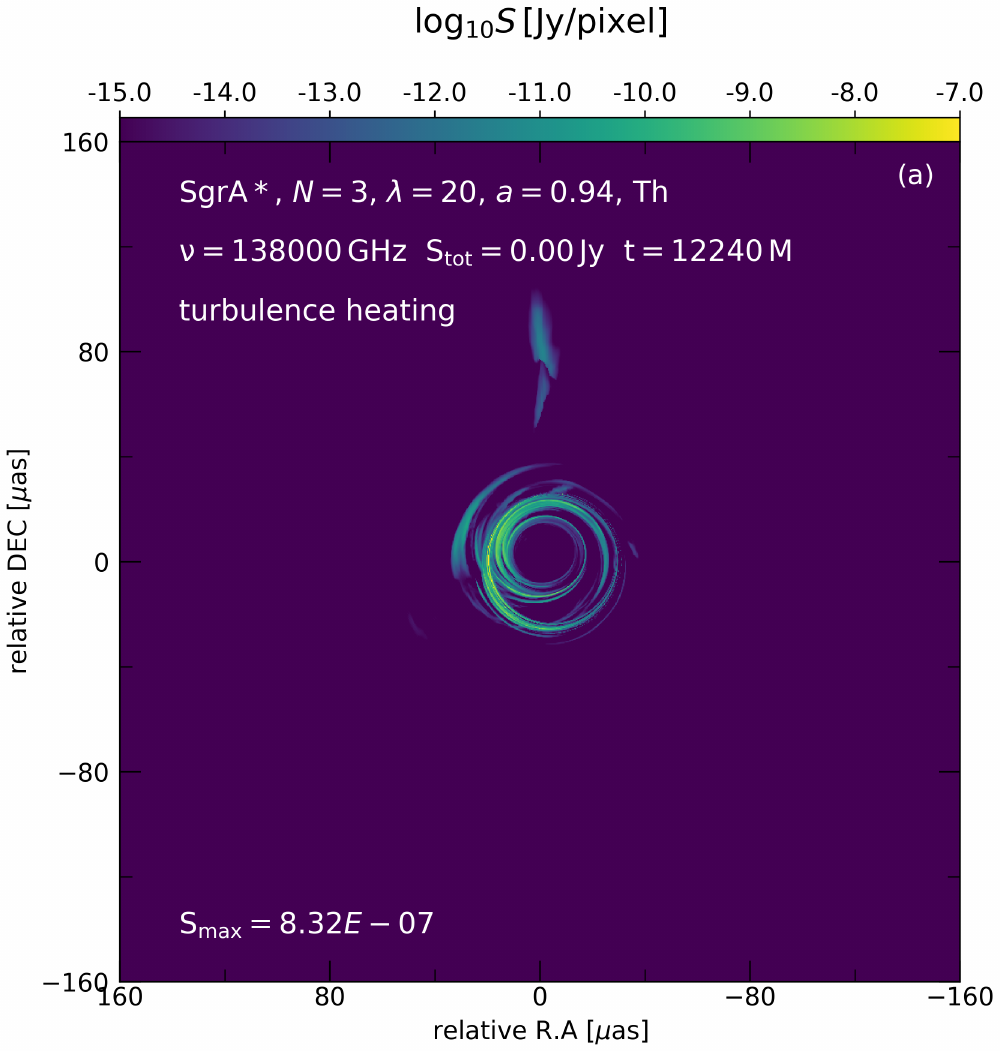}
	\includegraphics[width=.8\linewidth]{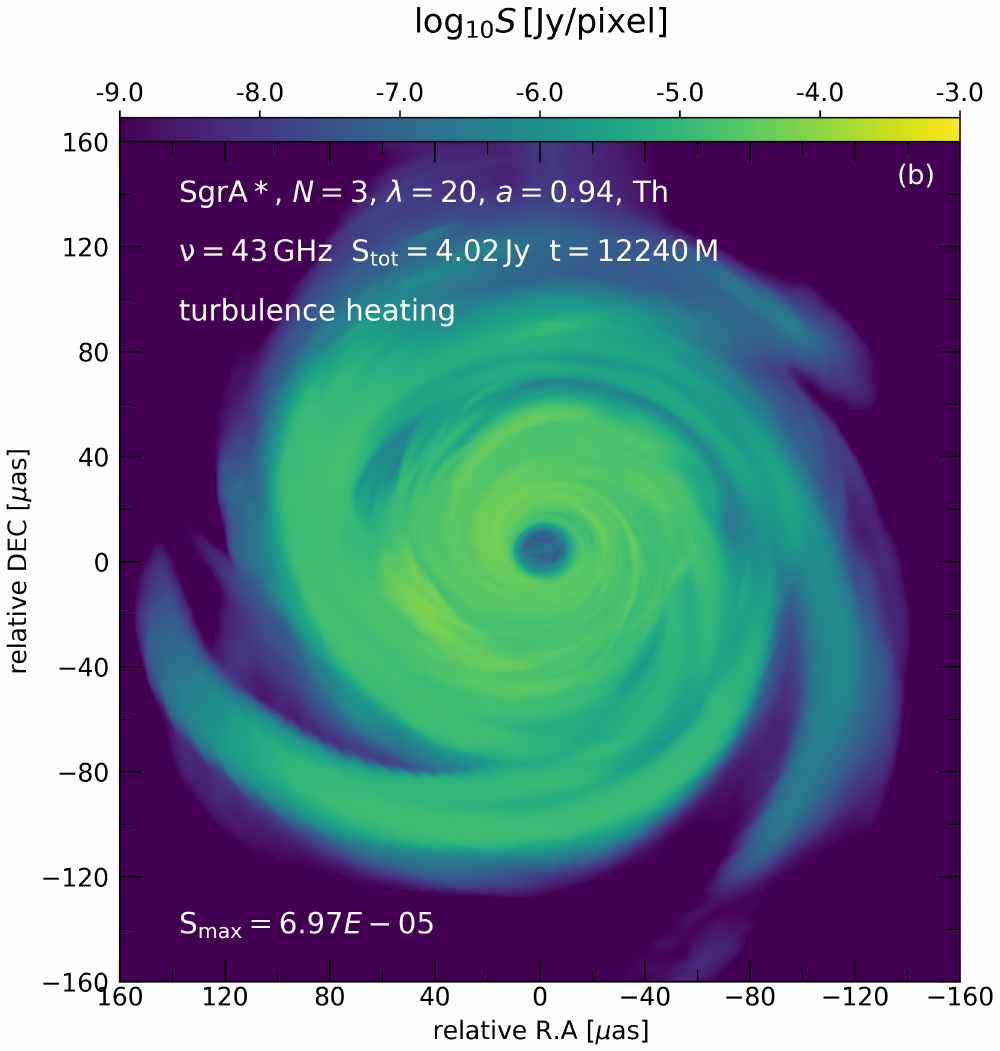}
    \caption{Panel a and b present the NIR and 43~GHz GRRT image from the turbulence model at $t=12,240\,\rm M$}
    \label{fig: 43GHz}
\end{figure}

To obtain the strength of the magnetic field inside flux ropes, we measure it at the regions with high electron temperature ($\Theta_{\rm e}>10$ obtained from turbulence heating model) and high density ($\rho>0.001$) of the case {\tt M20a3D} at $t=12,240\,\rm M$. The histogram of the magnetic field strength distribution is presented in Fig.~\ref{fig: B_distribution}. 
It indicates that inside the flux ropes, the magnetic field is relatively stronger than in other regions. In most of the regions, the magnetic field strength is $B=10^{-1.3}\approx 0.05$ in the simulation code unit. Applying the typical value of the magnetic field strength in the flux ropes, we present the relation between the emissivity under different observing frequencies and electron temperature in Fig.~\ref{fig: emissivity}.
At any frequency or magnetic field strength, emissivity and electron temperature are in positive correlation before the electron temperature reaches its peak value. After the peak, their relationship becomes negatively correlated. However, the position of the peak value of the emissivity depends on the observing frequency and local magnetic field strength. 

At 230 GHz, a stronger magnetic field makes the peak move to a lower electron temperature. It explains the suppression of the emission from the funnel region, which has a high electron temperature and high magnetic fields. The electron temperature of the filamentary structure seen in Fig.~\ref{fig: flare_Thetae} is about $\Theta_e=10\sim 100$, which corresponds to the high emissivity region at $230\,\rm GHz$ from Fig.~\ref{fig: emissivity} by using a typical magnetic field, $B=0.05$ in code unit. Accordingly, we see a clear filamentary structure in the GRRT images for the turbulence heating model (see Fig.~\ref{fig: Thetae}(a)) in the 230 GHz image.

Figure~\ref{fig: 43GHz} displays the GRRT images computed from the same snapshot used in Fig.~\ref{fig: flare_Thetae} at different frequencies (43~GHz and 134~THz). At a lower frequency, such as 43 GHz, it has a peak at a lower electron temperature (see the dotted line in Fig.~\ref{fig: emissivity}). Assuming optically thin emission, the lower frequency GRRT image has less filamentary structure due to the weaker emission from the high electron temperature flux ropes.
At higher frequencies, such as near-infrared (NIR) ($138\,\rm THz$), the peak of the emissivity coefficient occurs when $\Theta_{\rm e}>1,000$. Since the maximum electron temperature is usually $\sim 1,000$, which is always positively correlated with $\Theta_{\rm e}$ (see dashed line of Fig.~\ref{fig: emissivity}). As a result, the region with the highest electron temperature dominates the emission at $138\,\rm THz$. From the lower ($43\,\rm GHz$) to the higher frequencies (NIR), emission becomes more compact due to less contribution of outer accretion flows. It leads to an enhancement of the contribution of emission from flux ropes and shows more filament structure at a higher frequency.

\begin{figure}
    \centering
    \includegraphics[width=\linewidth]{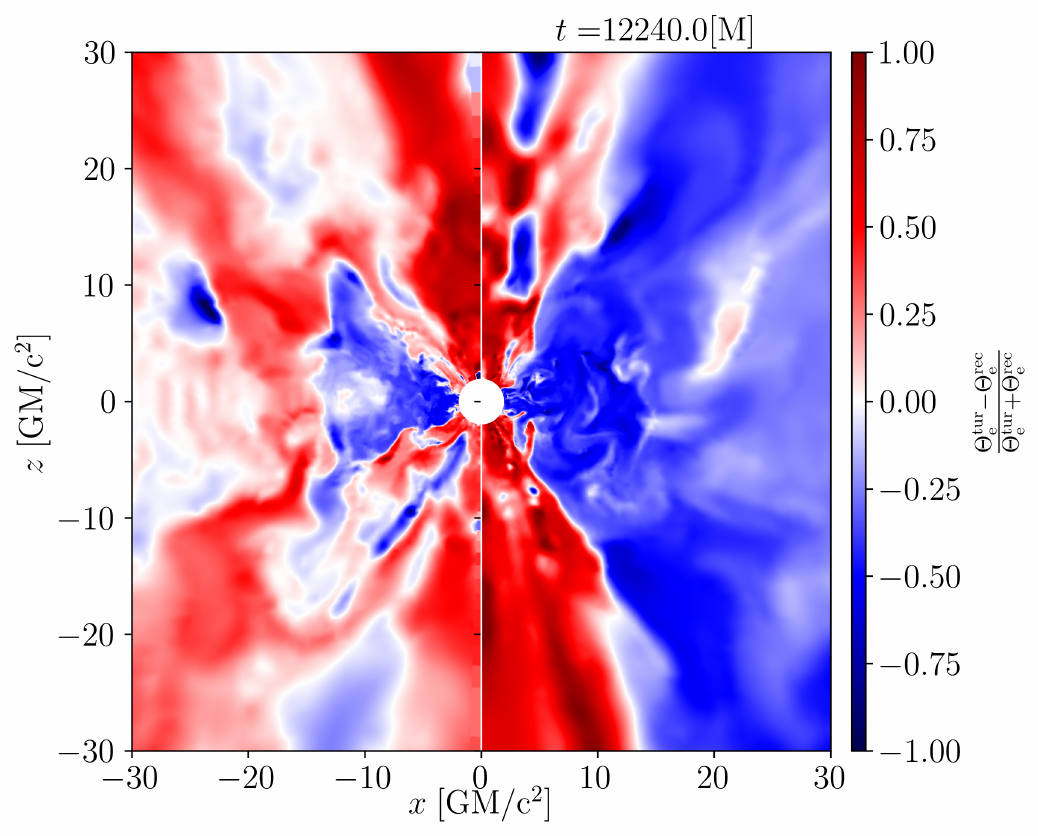}
    \caption{The distribution of the relative difference between the electron temperature distribution given by the turbulence model and reconnection model for the case {\tt M20a3D} at $t=12,240\,M$.} 
    \label{fig: eta}
\end{figure}

\begin{figure*}
\centering
	\includegraphics[width=.9\linewidth]{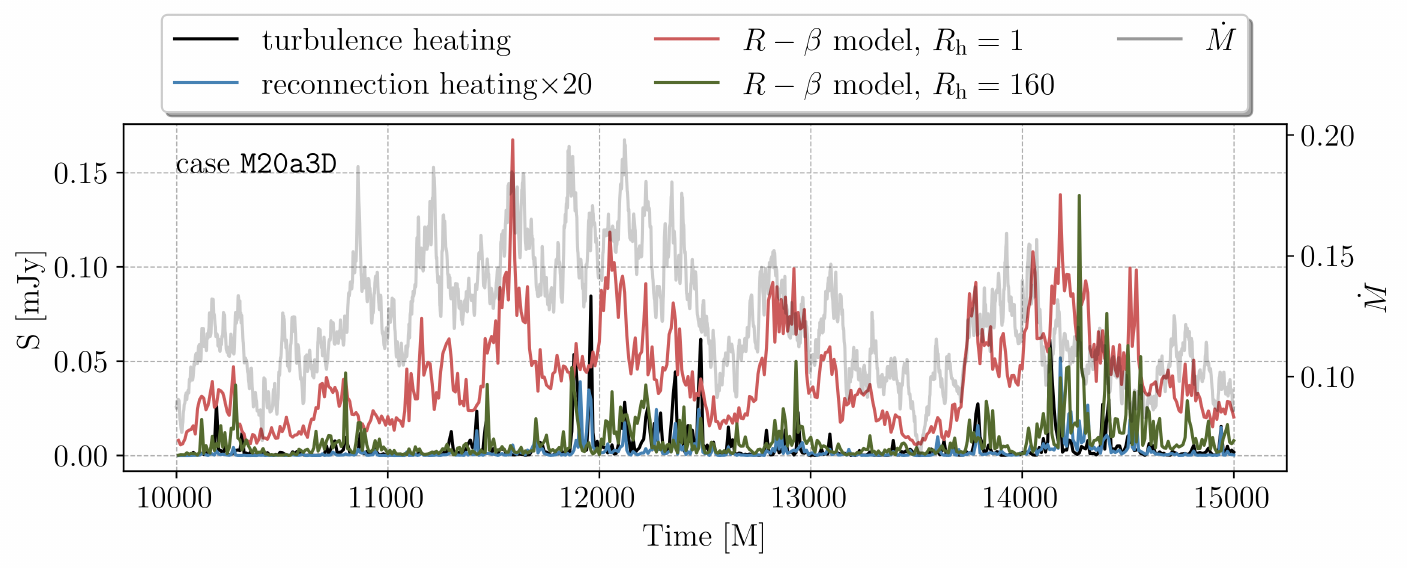}
    \caption{Same with Fig.~\ref{fig: flare_lc} but at NIR frequency.}
    \label{fig: NIR}
\end{figure*}

To understand the detailed differences between the emissions from the turbulence and reconnection heating models, we plot a pixel-by-pixel comparison of electron temperature between them for the case {\tt M20a3D} at $t=12,240\,\rm M$ in Fig.~\ref{fig: eta}. In the figure, we see the asymmetry between the left and right halves. It corresponds to the asymmetric structure of the accretion flow in the azimuthal direction related to the existence of the flux rope.
Figure~\ref{fig: eta} indicates that the turbulence heating model has a higher electron temperature in the jet and sheath than that in the reconnection heating model. On the other hand, the electron temperature for the reconnection heating model is higher than that of the turbulence heating model in the high-density disk region. Considering the similar mass accretion rates (turbulence model: $\dot{M}_{\rm tur}=2.19\times10^{-8}\,\rm M_{\odot}\,yr^{-1}$, reconnection model: $\dot{M}_{\rm rec}=1.19\times10^{-8}\,\rm M_{\odot}\,yr^{-1}$), the difference of $\Theta_{\rm e}$ is the key component for the significantly higher flux at $138\,\rm THz$ from the turbulence heating model than the reconnection heating model (see Fig.~\ref{fig: NIR}). The latter one has a lower maximum electron temperature inside the flux ropes. Although the torus electron temperature is higher in the reconnection model, it is still lower than the electron temperature of the flux ropes in the turbulence model.

\begin{figure*}
    \centering \includegraphics[height=0.37\linewidth]{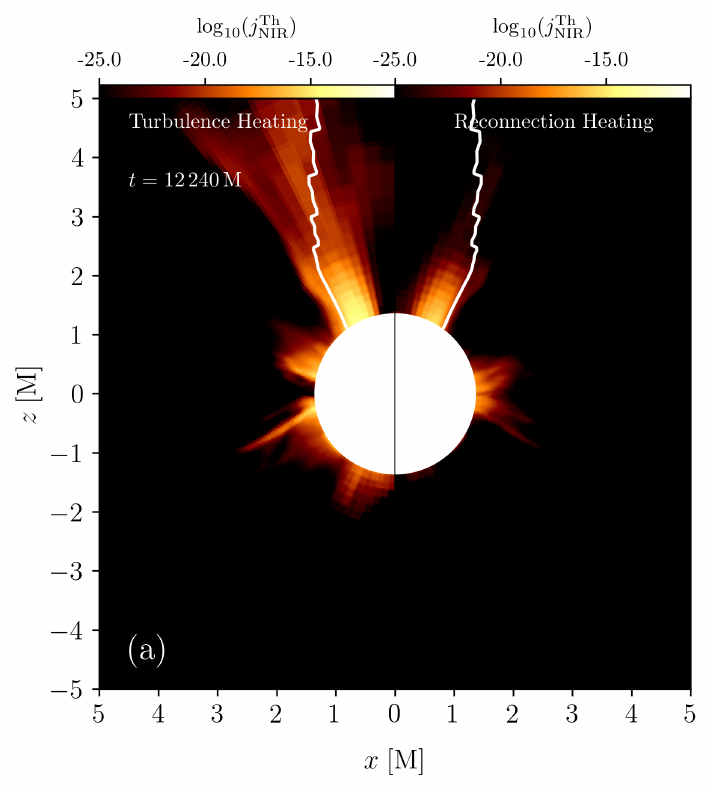}
\includegraphics[height=0.37\linewidth]{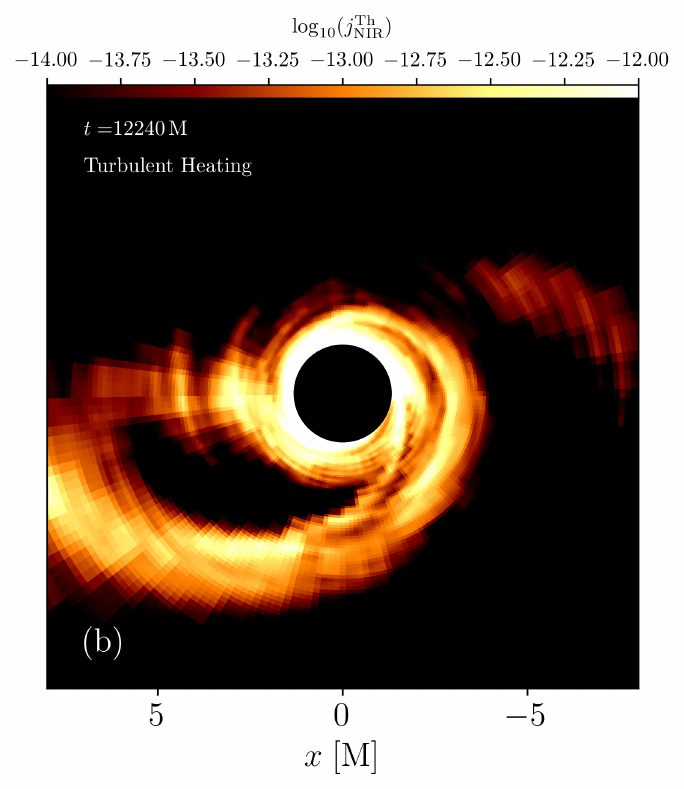}
\includegraphics[height=0.37\linewidth]{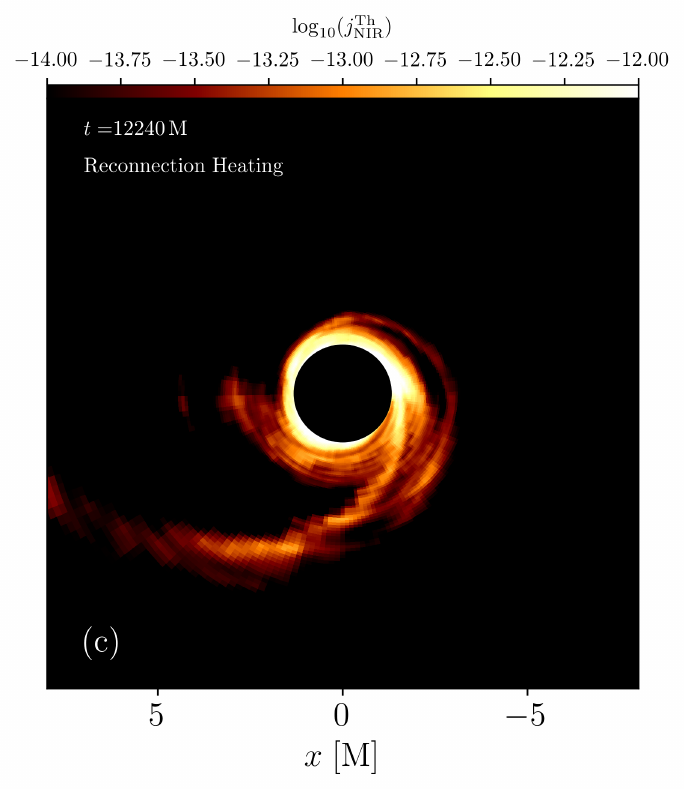}
    \caption{Panel (a) is the azimuthally averaged NIR emissivity distributions from turbulence ({\it left}) and reconnection ({\it right}) electron heating models; panel (b) and (c) are the zenith averaged NIR emissivity distributions of turbulence and reconnection heating model. }
    \label{fig:phi_avg}
\end{figure*}

\subsubsection{Emitting region at NIR frequency}

Since the mass accretion rates of different models are calibrated at 230 GHz, the average near-infrared flux varies among the models. The light curves of the case $\tt M20a3D$ with different electron heating models are presented in Fig.~\ref{fig: NIR}\footnote{Due to the significantly lower NIR luminosity of the reconnection heating model, we plot it multiplying by a factor of 20.}. Unlike the situation at $230\,\rm GHz$, the two-temperature models have less similarity with the parameterized $\rm R-\beta$ models. The reconnection heating model has a significantly lower flux (about a magnitude) than other models, while the R-$\beta$ model with $R_{\rm h} = 1$ has the highest flux. The turbulence heating model still shows some similarity with the R-$\beta$ model with $R_{\rm h}=160$.
Focusing on the two periods with frequent flaring activities at $t=12,000-13,000\,\rm M$ and $t=14,000-15,000\,\rm M$ in Fig.~\ref{fig: NIR}, the light curve of the turbulence heating model shows more spiked structure than that in the $\rm R-\beta$ model with both $R_{\rm h}$ values as discussed here. Less variation but stronger emission is seen in the R-$\beta$ model with $R_{\rm h}=1$. Such light curves in the four different models are caused by the different emitting regions at the NIR frequency.
Since the emission at NIR frequencies is dominated by non-thermal emission, the peaks of the light curves observed in Fig.~\ref{fig: NIR} are one to two orders of magnitude lower than the observed value for Sgr\,A$^{*}$(5-25\,\rm mJy) \citep{Abuter2020}. Accordingly, consideration of non-thermal emission is required to explain observed NIR emission \citep{2006AAS...20911207Y, dodds-eden_time-dependent_2010, witzel_rapid_2021, zhao_impact_2023}. We will investigate it in our future work. 

Although the non-thermal component is not included in this work, the thermal NIR emission property is also interesting. The NIR light curves are strongly related to the regions with very high electron temperatures ($\Theta_{\rm e}\gtrsim 100$). The maximum electron temperature is roughly $\Theta_{\rm e}\sim 500$ in all models.
According to the relationship in Fig.~\ref{fig: emissivity}, with fixed magnetic field strength, the thermal-synchrotron emission increases with electron temperature until $\Theta_{\rm e}\lesssim 1,000$. Therefore, at the NIR frequency, the thermal synchrotron emission mainly comes from the high electron temperature regions.
As \cite{stad1106} suggested, the R-$\beta$ model is not able to reflect the detailed structure of the high-temperature plasmoids and current sheet. In this 3D simulation result, a more detailed structure is also revealed by two-temperature electron heating models (see Fig.~\ref{fig: Thetae}).
To determine the emitting region at NIR frequency, we plot the NIR emissivity distributions of turbulence and reconnection models in averaged $\phi$ and $\theta$ directions, respectively, at $t=12,240\,\rm M$ in Fig.~\ref{fig:phi_avg}.
In the turbulence heating model, NIR emission mainly comes from the sheath region. On the other hand, in the reconnection heating model, the weaker NIR emission originates mainly from the torus.
Comparing with Fig.~\ref{fig:phi_avg}(b) and (c), we see a bright spiral-like structure in the $\theta$-averaged figures, which is contributed by the flux rope. The turbulence model generates a higher and more extended high emissivity region than that of the reconnection model leading to higher flux.

Since the flux ropes have an important contribution at NIR frequency, which can be a potential source for the observed NIR flares \citep[e.g.][]{Abuter2023}. A low-inclination orbital plane is suggested for the 3D orbit of the flares \citep{Levis}.
It conflicts with the MAD model that the orbital motion of the flares located in the equatorial plane (inclination angle $\sim 90^\circ$) \citep[e.g.][]{Scepi2022,2020MNRAS.497.4999D,porth_flares_2021}.
From our simulations, we see that most flux ropes are located in the sheath region with a relatively small inclination angle $\sim 30^\circ$ and are highly variable. The orbiting time scale of the flux ropes is $\sim 50\,\rm M$. In most cases, they can only last roughly one orbital period before they become too weak to be observed.
Therefore, the duration of the NIR flares is also roughly $10-100\,\rm M$, as seen in Fig.~\ref{fig:phi_avg}. The frequent energy release in these regions gives rise to the highly variable light curve in the turbulence heating model. Because the dominant emitting region for this model is the flux ropes (see Fig.~\ref{fig:phi_avg}(b)). Thus, the dynamics of the high-temperature plasma directly influence the behavior of light curves.

 

\section{Conclusions}

We have performed 3D, two-temperature GRMHD simulations of magnetized accretion flows onto a rotating black hole with multiple magnetic loops. We use two different initial magnetic loop sizes ($\lambda=20$, $80$). For electron heating prescription, we apply turbulence and reconnection heating models. The parameterized R-$\beta$ model is also included for comparison with the two-temperature electron heating models. The summaries of our results are presented below.

\begin{enumerate}
    \item The simulations show that the initial magnetic loop size affects the strength of the magnetic dissipation and the MRI inside the torus. The smaller loop size results in stronger dissipation and weaker MRI, while the larger loop size leads to a partial transition from SANE to MAD flow. This finding is consistent with our previous 2D simulations~\citep{stad1106}. However, in 3D, the transition is not fully observed due to the insufficient magnetic field to quench the accretion.
    \item When the first loop reaches the horizon, the flow resembles that of the simulations with a single loop seen in \cite{2021MNRAS.506..741M}. The light curves at 230~GHz from different electron heating models are also similar, regardless of whether they use two-temperature GRMHD or parameterized R-$\beta$ models. However, when the different polarities of a magnetic loop  accrete, a large amount of reconnection happens, and a highly turbulent accretion flow is generated.
    \item After the accretion flow is fully turbulent, the light curves from the different electron heating models show distinct differences. The turbulence heating model and the R-$\beta$ model with $R_{\rm h}=160$ show more variability than the reconnection heating model and the $R_{\rm h}=1$ one. Because the former two models have strong emissions from the sheath region, where the accretion flow is more variable due to the frequent reconnection events. On the other hand, the reconnection heating model and the R-$\beta$ model with $R_{\rm h}=1$ have more emission from the equatorial plane with less turbulence. Thus, the variability of light curves from different electron heating models in the multi-loop simulations reflects the dynamics of the emitting regions.
    \item Electron temperature has a direct impact on thermal emissivity, which is also related to observing frequency. At higher frequencies, most of the emissions come from regions with high electron temperatures, such as erupting flux ropes. The flux ropes have both high density and electron temperature, and they are more variable than the accretion flow in the equatorial plane. Therefore, they are responsible for the filamentary structure seen in GRRT images and the light curve variability at high frequencies, such as NIR. However, at lower frequencies, e.g., 43~GHz, a less distinct filamentary structure is observed than that at 230~GHz and NIR.
    \item In comparison with the two-electron heating models, the turbulence heating one shows more variance in the distribution of electron temperature. It provides a higher electron temperature in the jet and flux ropes but a lower temperature than the reconnection one in the accretion disk. This result is also observed in our 2D simulations \citep{stad1106}. The global distribution of electron temperature still matches the R-$\beta$ model. However, the difference comes from small-scale regions. 
\end{enumerate}

This paper focuses on the 3D GRMHD dynamics and thermal synchrotron emission from the multi-loop magnetic field configuration. 
The accretion flow from multiple magnetic loop configurations does not show clear features of the MAD state, which is more favored for Sgr~A$^*$ suggested by EHT observation \citep{event_horizon_telescope_collaboration_first_2022}. However, none of the models (irrespective of SANE or MAD) pass all the constraints from EHT observations. Most of them failed to satisfy the variability constraint. Considering the flaring activities of Sgr~A$^*$, we try to use the accretion flow generated from a multi-loop magnetic configuration to explain it. We see reconnection events in the jet sheath, which we find related to the observed flares. However, with only thermal emission, the magnitude is not strong enough to compare with NIR observations. A study considering non-thermal emissions from flux ropes will be implemented in our next work, and we plan to investigate its contribution to the NIR flux. Another possible reason for the underestimated NIR emission may originate from the limited resolution of our 3D simulations. Because of the low resolution, tearing instability as well as plasmoids are not well resolved. It may underestimate the electron temperature in these regions. It indicates that higher-resolution simulations are required to resolve the flaring events of Sgr~A$^*$. More comparisons with observations such as visibility amplitude morphology, M-ring fits, etc. are required to determine the applicability of the simulations of multi-loop configurations to the Sgr~A$^*$.

In this work, flux ropes with relatively ordered magnetic fields are observed and proved to contribute to the total flux when the observing frequencies increase. It suggests that relatively higher polarized emission will be observed from these flux ropes. We will investigate it via polarized GRRT calculations in our future work.

\begin{acknowledgements}
The authors gratefully acknowledge insightful discussions with Dr. Xi Lin from Shanghai Astronomical Observatory.
This research is supported by the National Natural Science Foundation of China (Grant No.\,12273022), the Shanghai Municipality orientation program of Basic Research for International Scientists (Grant No.\,22JC1410600), and the National Key R\&D Program of China (No.\,2023YFE0101200). ZY acknowledges support from a UKRI Stephen Hawking Fellowship. CMF is supported by the DFG research grant ``Jet physics on horizon scales and beyond" (Grant No. 443220636).
The simulations were performed on TDLI-Astro, Pi2.0, and Siyuan Mark-I at Shanghai Jiao Tong University.
\end{acknowledgements}

%

\begin{appendix}

\section{Electron Temperature Profile}

\begin{figure}
    \centering
    \includegraphics[width=\linewidth]{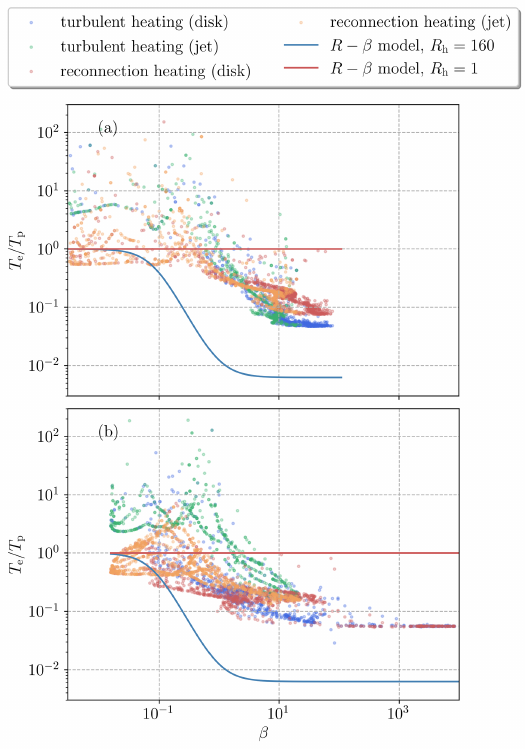}
    \caption{$T_{\rm e}/T_{\rm p}-\beta$ diagram of the average of collapse data from $15,000-20,000\,\rm M$ of case {\tt M80a3D} (a) and $10,000-15,000\,\rm M$ of case {\tt M20a3D} (b). Different colors mark the jet and disk components and the turbulence/reconnection heating models. Jet and disk are divided with criteria of magnetization $\sigma>1$. The red and blue lines represent the R-$\beta$ model with $R_{\rm h}=1, 160$ respectively.}
    \label{fig: TeTp-beta}
\end{figure}
The electron distribution of the two-temperature models implies a more complicated relationship between $T_{\rm e}/T_{\rm p}$ and plasma $\beta$. Therefore, we plot the $T_{\rm e}/T_{\rm p}-\beta$ relations from the time and azimuthal averaged data of the cases {\tt M80a3D} and {\tt M20a3D} in Fig.~\ref{fig: TeTp-beta}. In the jet regions of both two-temperature models, the existence of points with $T_{\rm e}/T_{\rm p} > 1$ indicates that electron temperature is dominant in some regions. The $T_{\rm e}/T_{\rm p}$ ratio scatters at the low $\beta$ region, implying the simple R-$\beta$ model may not describe the underlying physics in this region. 
For the weak magnetic field region, i.e., high plasma $\beta$, both two-temperature models converge to the same value in case {\tt M20a3D} (see the lower panel of Fig.~\ref{fig: TeTp-beta}). However, at $\beta\sim 10$, higher $T_{\rm e}/T_{\rm p}$ is observed in the reconnection heating model, corresponding to the bluish part of the disk region of Fig.~\ref{fig: eta} locates. A similar phenomenon has been observed in our previous 2D simulations (see \cite{stad1106}). 
\end{appendix}

\end{document}